\documentclass[]{spie}  
\usepackage[]{graphicx}

\usepackage{graphicx}
\usepackage{xspace}
\usepackage{url}
\usepackage{eurosym}
\usepackage[polutonikogreek,english]{babel}

\graphicspath{{./Figures/}}
\newcommand{\uas}{\ensuremath{\mu\mbox{as}}\xspace}
\newcommand{\microns}{\ensuremath{\mu\mbox{m}}\xspace}
\newcommand{\meters}{\ensuremath{\mbox{m}}\xspace}
\newcommand{\AU}{\ensuremath{\mbox{AU}}\xspace}
\newcommand{\pc}{\ensuremath{\mbox{pc}}\xspace}

\newcommand{\Msun}{\ensuremath{\mbox{M}_\odot}\xspace}

\newcommand{\MEarth}{\ensuremath{\mbox{M}_\oplus}\xspace}
\newcommand{\SNR}{\ensuremath{\mbox{SNR}}\xspace}

\newcommand{\Nvisits}{\ensuremath{N_{\rm visits}}\xspace}

\newcommand{\Tvisits}{\ensuremath{T_{\rm visit}}\xspace}

\newcommand{\Mstar}{\ensuremath{M_\ast}\xspace}
\newcommand{\Mplanet}{\ensuremath{M_P}\xspace}

\urldef{\neaturl}\url{http://neat.obs.ujf-grenoble.fr}
\urldef{\neattargetsurl}\url{http://neat.obs.ujf-grenoble.fr/IMG/xls/Proposal_targets_total.xls}

\title{NEAT: a space born astrometric mission for the detection and characterization of nearby habitable planetary systems}

\author{Fabien Malbet\supit{a}, Renaud Goullioud\supit{b},
  Pierre-Olivier Lagage\supit{c}, Alain L\'eger\supit{d}, Mike
  Shao\supit{b}, Antoine Crouzier\supit{a}, and the NEAT consortium\supit{e}
\skiplinehalf
\supit{a}UJF-Grenoble 1 / CNRS-INSU, Institut de Plan\'etologie  et d'Astrophysique de Grenoble (IPAG), UMR 5274, Grenoble, France \\
\supit{b}Jet Propulsion Laboratory (JPL), California Institute of Technology, Pasadena, USA\\
\supit{c}Laboratoire AIM, CEA-IRFU / CNRS-INSU / Universit\'e Paris Diderot, CEA Saclay, Gif-sur-Yvette Cedex, France\\
\supit{d}Universit\'e Paris Sud / CNRS-INSU, Institut d'Astrophysique Spatiale (IAS) UMR 8617, Orsay, France\\
\supit{e}Full list of NEAT proposal members at \texttt{http://neat.obs.ujf-grenoble.fr}\\
}

\authorinfo{Further author information: (Send correspondence to Fabien.Malbet@obs.ujf-grenoble.fr)}

\begin{document} 
  \maketitle 

\begin{abstract}
The NEAT (Nearby Earth Astrometric Telescope) mission is a proposal submitted to ESA for its 2010 call for M-size mission within the Cosmic Vision 2015-2025 plan. The main scientific goal of the NEAT mission is to detect and characterize planetary systems in an exhaustive way down to 1 Earth mass in the habitable zone and further away, around nearby stars for F, G, and K spectral types. This survey would provide the actual planetary masses, the full characterization of the orbits including their inclination, for all the components of the planetary system down to that mass limit. NEAT will continue the work performed by Hipparcos and Gaia by reaching a precision that is improved by two orders of magnitude on pointed targets. 
\end{abstract}



\section{Introduction}
\label{sec:introduction}

Exoplanet research has grown explosively in the past decade, supported
by improvements in observational techniques that have led to
increasingly sensitive detection and characterization. Among many
results, we have learned that planets are common, but their physical
and orbital properties are much more diverse than originally
thought. 

A lasting challenge is the detection and characterization of planetary
systems consisting in a mixed cortege of telluric and giant planets,
with a special regard to telluric planets orbiting in the habitable zone
(HZ) of Sun-like stars. The accomplishment of this goal requires the
development of a new generation of facilities, due to the intrinsic
difficulty of detecting Earth-like planets with existing instruments.
The proposed NEAT mission has been designed to enter a
new phase in exoplanetary science by delivering an enhanced capability
of detecting small planets in the Habitable Zone

In Sect.~\ref{sec:neat-science}, we present the science objectives of
NEAT, we describe the principle of the differential astrometry
technique and we give a list of potential targets. In
Sect.~\ref{sec:neat-concept}, after listing the technical challenges,
we present the instrumental concept. We explain how to reach the
performance and we give a summarized description of the payload, the
mission and the spacecraft. In Sect.~\ref{sec:discussion}, we discuss
both astrophysical and technical issues. Recommandations by the
community summarized in Sect.~\ref{sec:perspectives} is an incentive
to pursue the development of this mission in the future.

\section{Science objectives}
\label{sec:neat-science}

\subsection{Main science questions}
\label{sec:science-cases}

The prime goal of NEAT is to detect and characterize planetary systems
orbiting bright stars in the solar neighborhood that have a planetary
architecture like that of our Solar System or an alternative planetary
system made of Earth mass planets. It will allow the detection around
nearby stars of planets equivalent to Venus, Earth, (Mars), Jupiter,
and Saturn, with orbits possibly similar to those in our Solar System.
It will permit to detect and characterize the orbits and the masses of
many alternate configurations, e.g.\ where the asteroid belt is
occupied by another Earth mass planet and no Jupiter. The NEAT mission
will answer the following questions:
\begin{itemize}
  \item What are the dynamical interactions between giant and
    telluric planets in a large variety of systems?
  \item What are the detailed processes involved in planet formation as
    revealed by their present configuration?
  \item What are the distributions of architectures of planetary systems
    in our neighborhood up to $\approx15$\,pc?
  \item What are the masses, and orbital parameters, of telluric planets that are
    candidates for future direct detection and spectroscopic characterization
    missions?
\end{itemize}
Special emphasis will be put on planets in the \emph{Habitable Zone}
because this is a region of prime interest for astrobiology. Indeed
orbital parameters obtained with NEAT will allow spectroscopic
follow-up observations to be scheduled precisely when the
configuration is the most favorable.

\subsection{High-precision differential astrometry}
\label{sec:astrometry}

\begin{figure}[t]
  \centering
  \includegraphics[width=0.4\hsize]{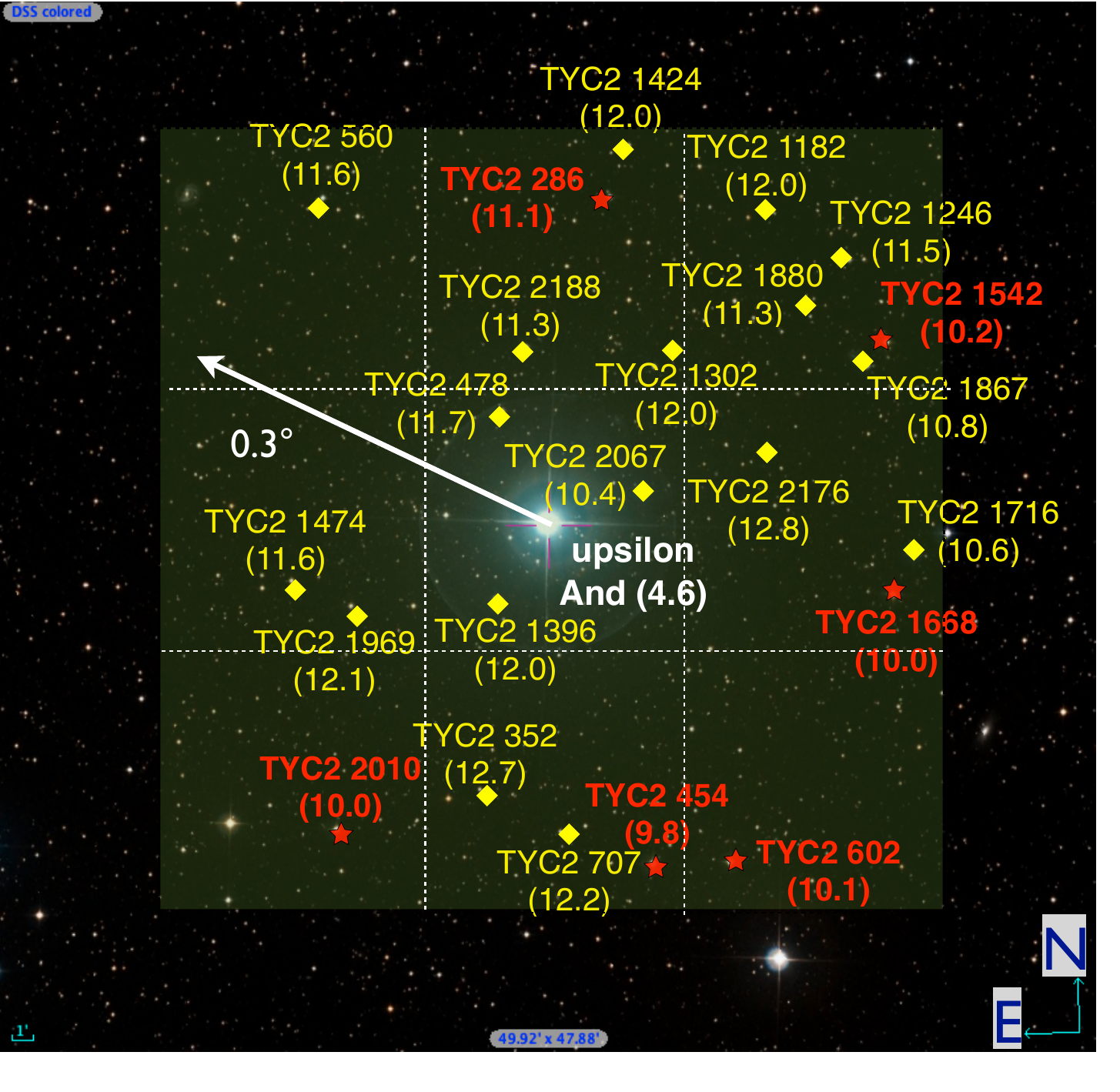}
  \caption{$0.3^\circ$ stellar field around upsilon Andromedae, a
    proposed NEAT target. There are six possible reference stars in
    this field marked in red (five $V<11$ stars
    and a $V=11.1$ one).}
  \label{fig:ups-and}
\end{figure}

\begin{figure*}[t]
  \centering
  \begin{tabular}{ccc}
      \includegraphics[width=0.3\hsize]{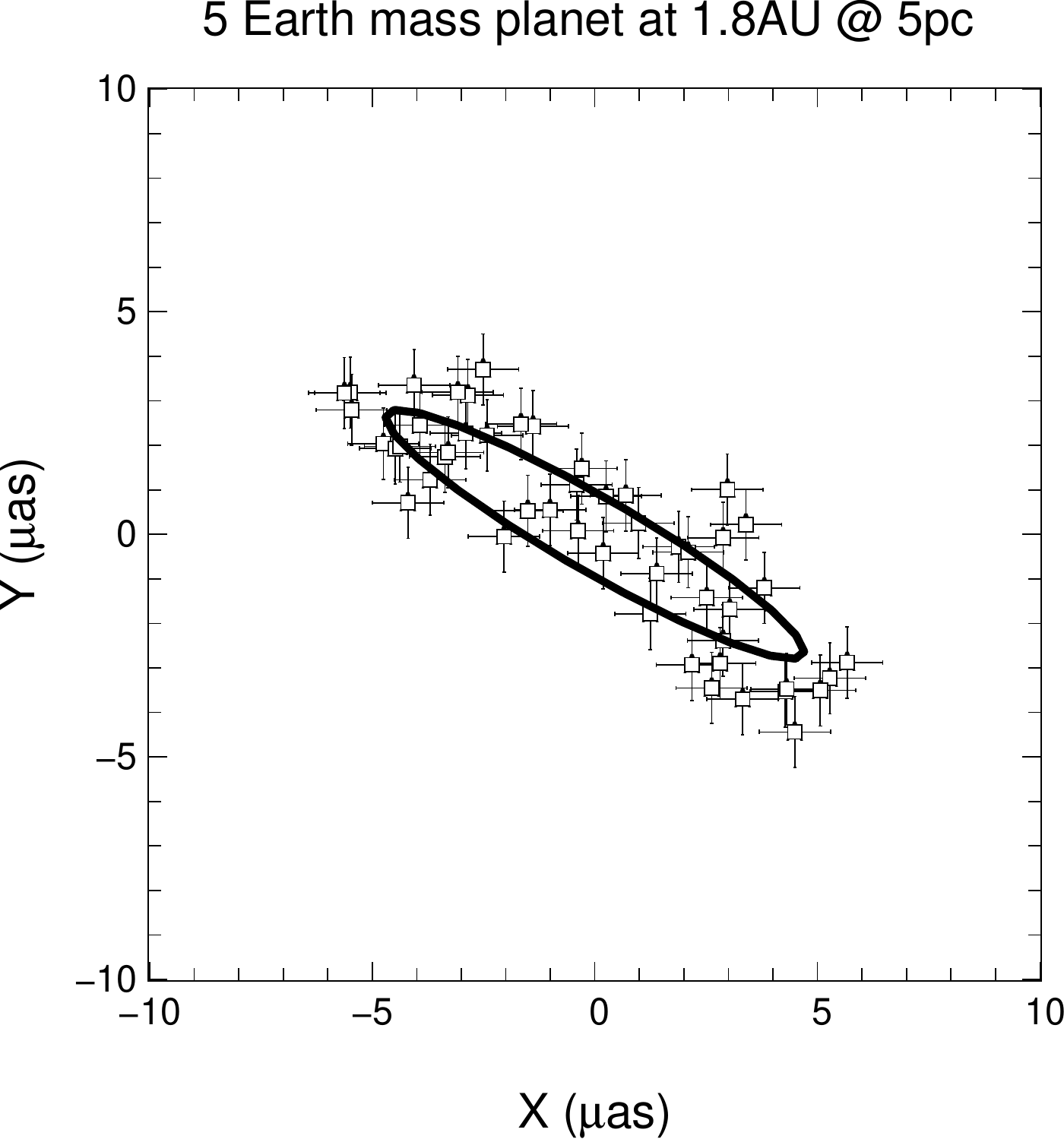}&
      \includegraphics[width=0.3\hsize]{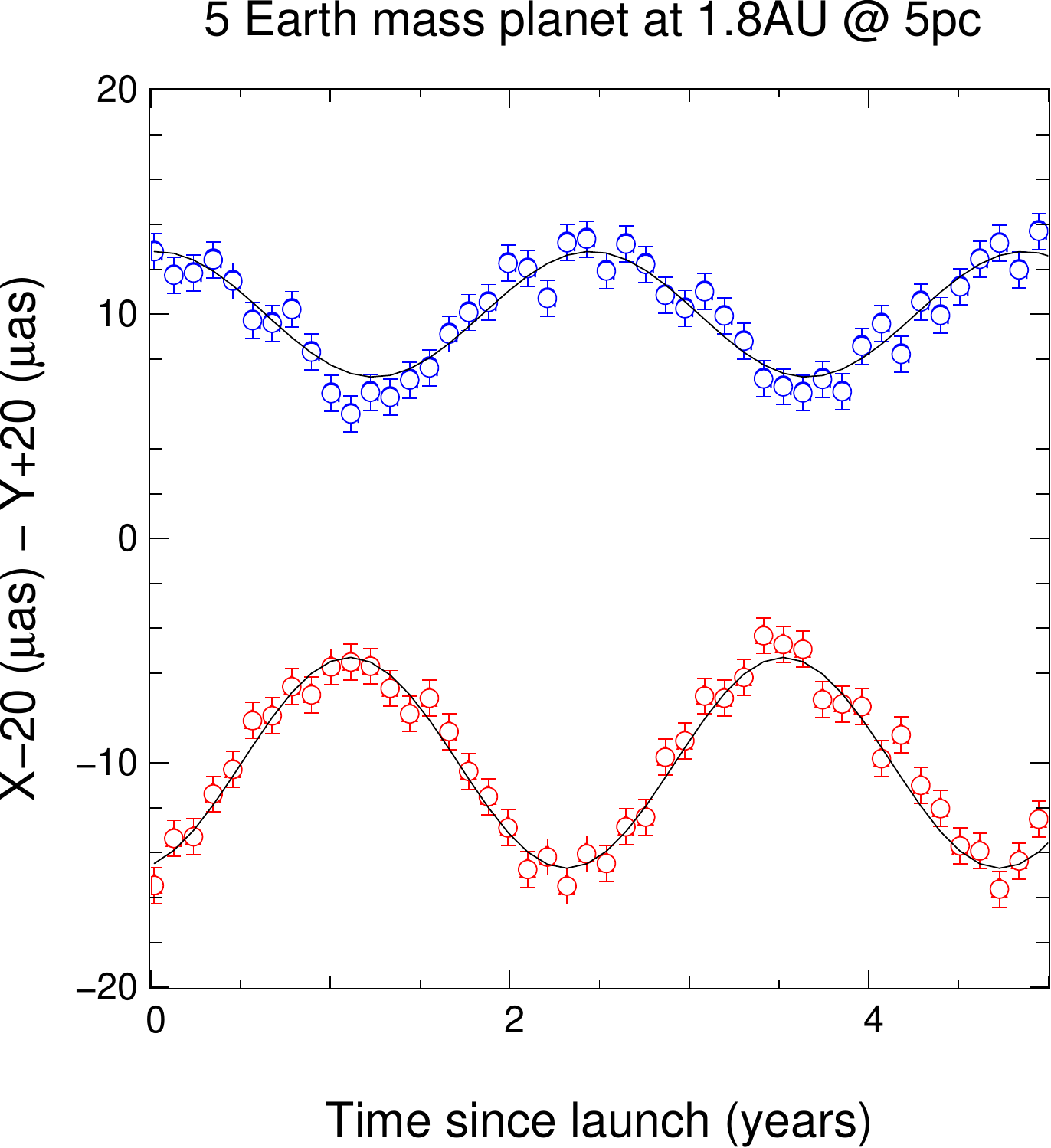}&
      \includegraphics[width=0.3\hsize]{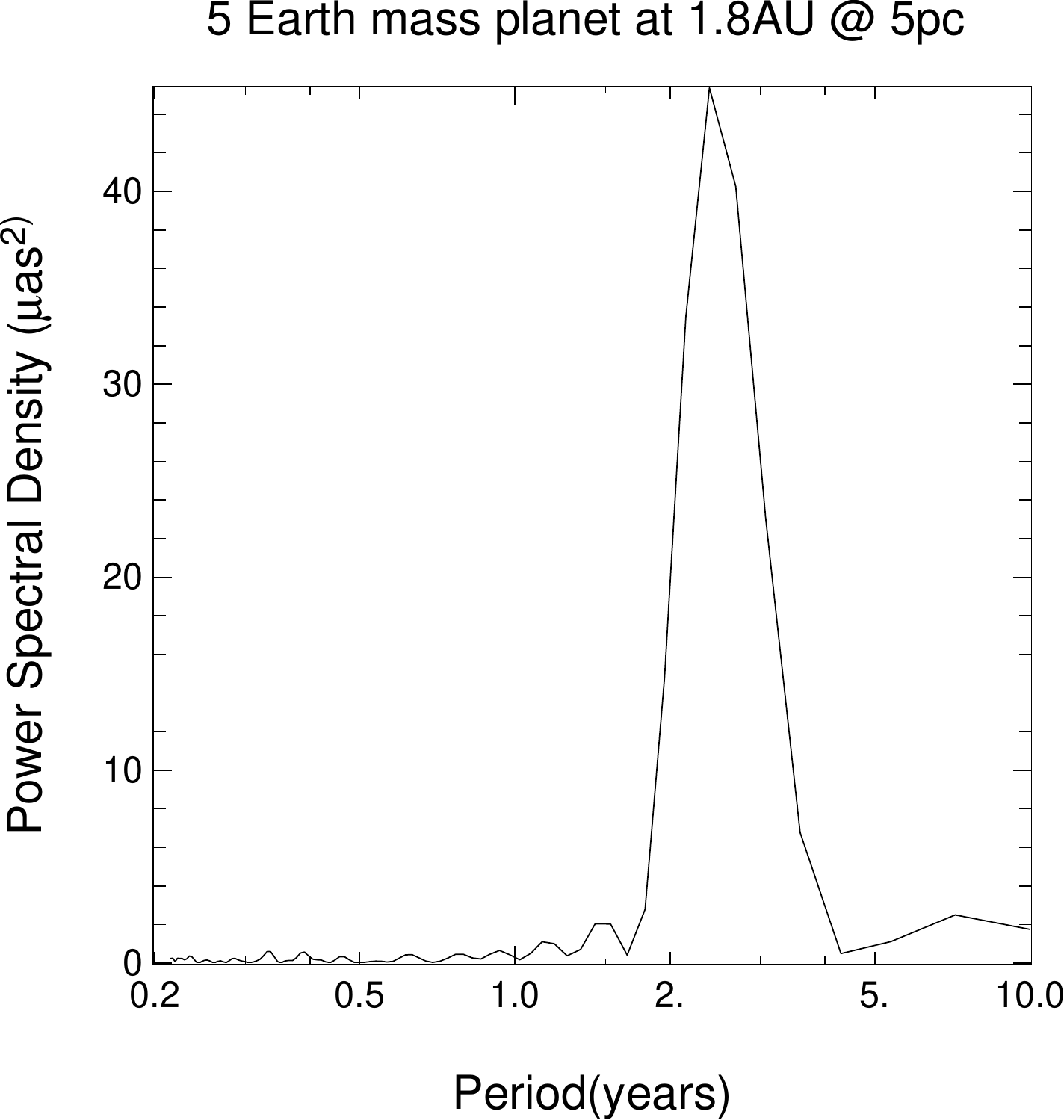}\\
      \includegraphics[width=0.3\hsize]{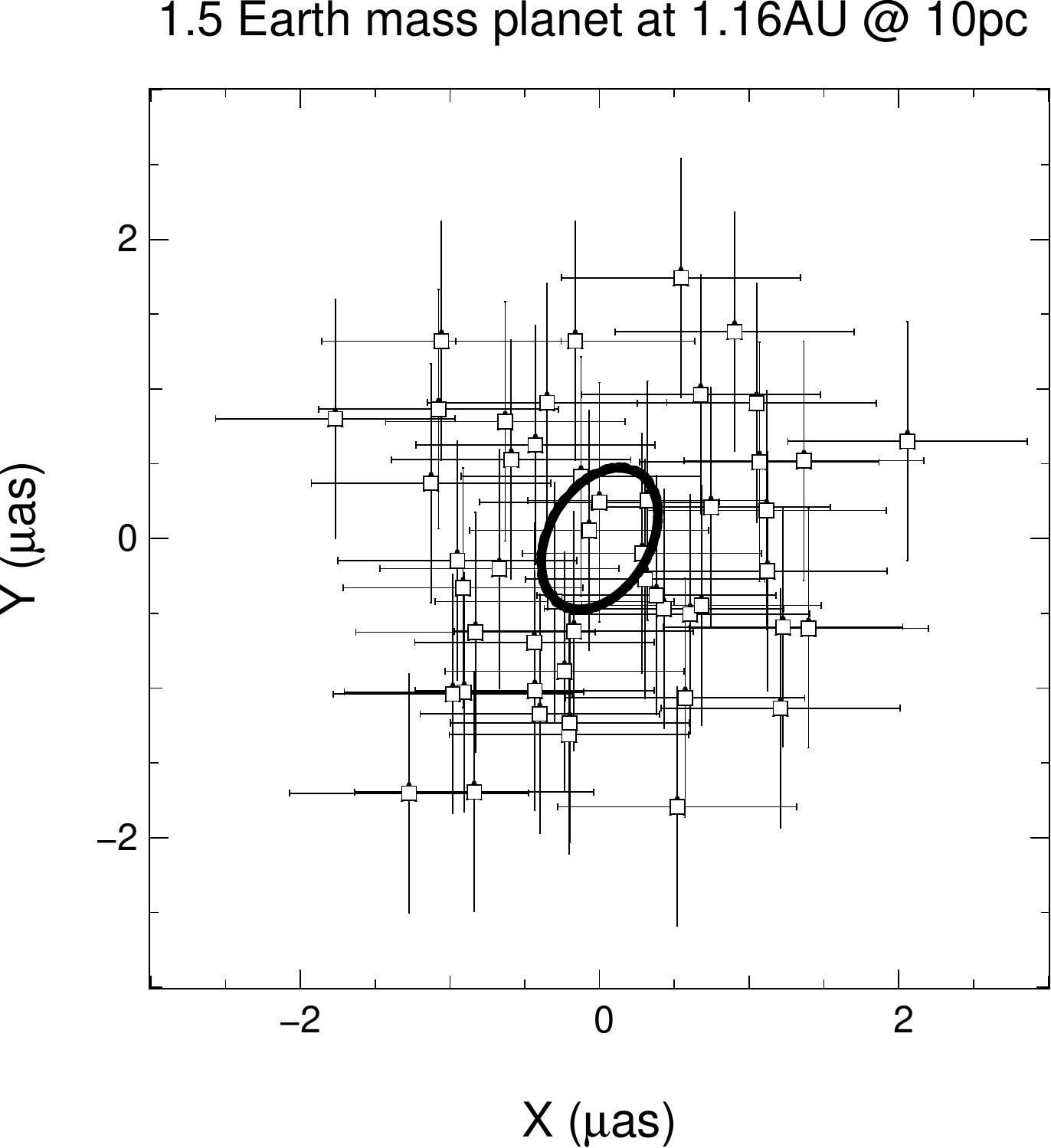}&
      \includegraphics[width=0.3\hsize]{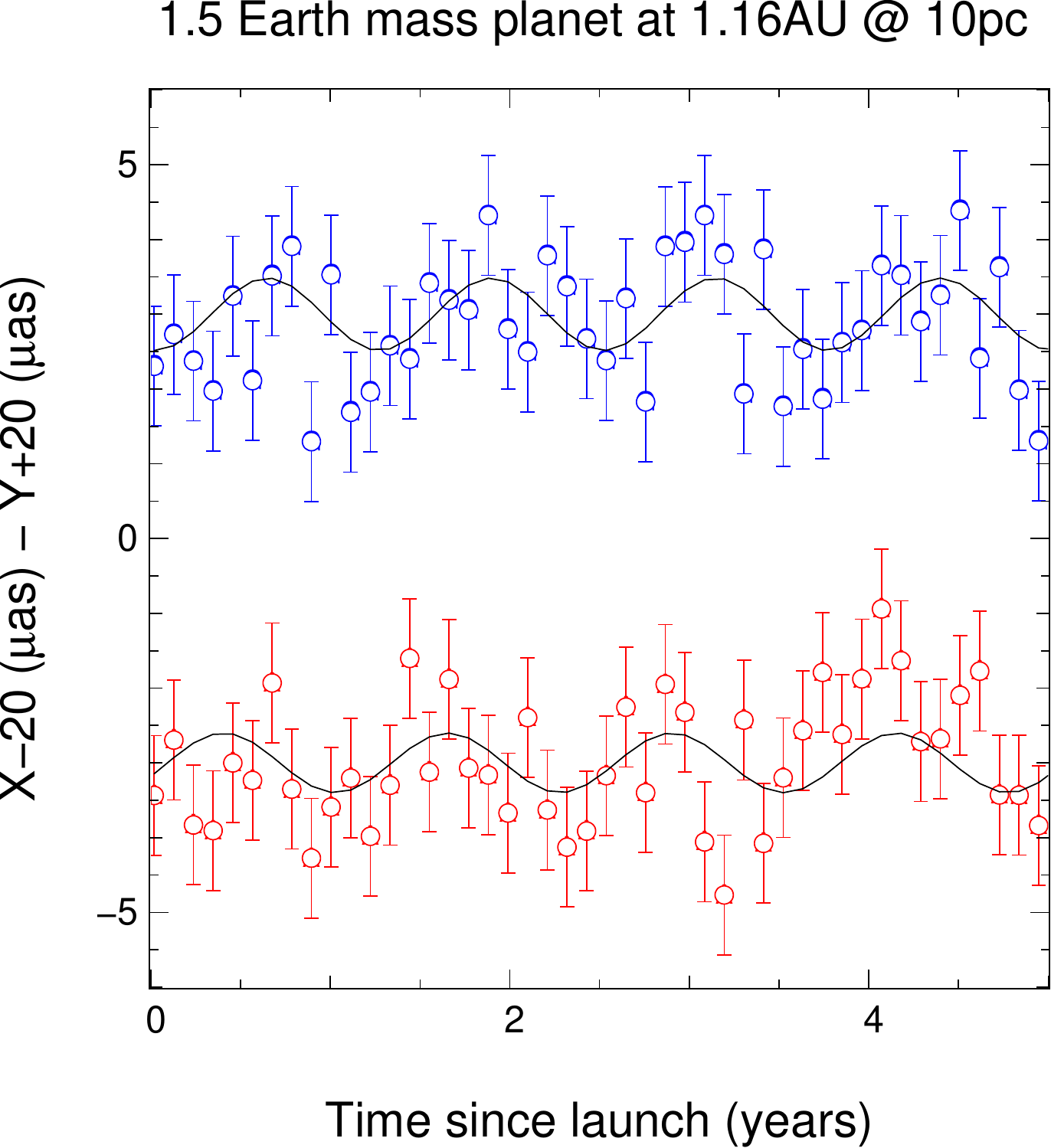}&
      \includegraphics[width=0.3\hsize]{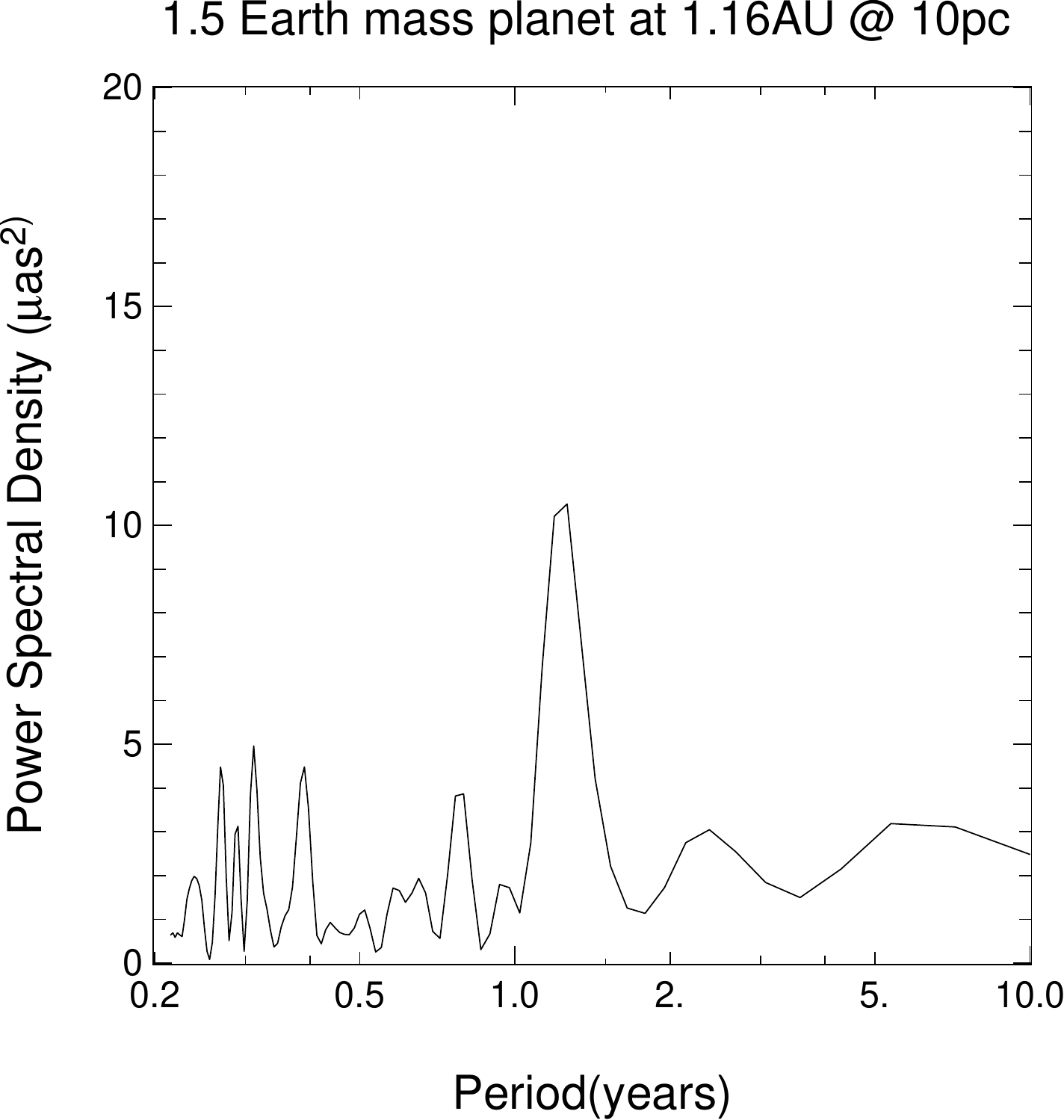}\\
  \end{tabular}
  \caption{Simulation of astrometric detection of two type of planets
    with 50 NEAT measurements (RA and DEC) over 5 yrs. Top row
    parameters corresponds to an easy detection:
    $\Mplanet=5\,\MEarth$, $a=1.8\,AU$, $\Mstar=1\,\Msun$,
    $D=5\,\pc$. Bottom row parameters corresponds to a more
    challenging detection: $\Mplanet=1.5\,\MEarth$, $a=1.16\,AU$,
    $\Mstar=1\,\Msun$, $D=10\,\pc$.  Left panels: sky plot showing the
    astrometric orbit (solid curve) and the NEAT measurements with
    error bars; middle panels: same data but shown as time series of
    the RA and DEC astrometric signal (an offset has been put to
    separate them); right panels: joint Lomb-Scargle periodogram for right
    ascensions and declinations simultaneously. All measurements
    corresponds to 1\,h visits with a 0.8\,\uas precision. This
    precision can be improved or degraded by a longer or a shorter
    visit time.}
    \label{fig:earth-detection}
\end{figure*}

The principle of NEAT is to measure accurately the offset angles
between a target and 6-8 distant reference stars with the aim of
differentially detecting the reflex motion of the target star due to the
presence of its planets. An example of a field that will be
observed is shown in Fig.~\ref{fig:ups-and} and a simulation of what
will be measured is displayed in Fig.~\ref{fig:earth-detection}.

The output of the analysis is a \emph{comprehensive} determination of
the mass, orbit, and ephemeris of the different planets of the
\emph{multiplanetary system} (namely the 7 parameters $\Mplanet$, $P$,
$T$, $e$, $i$, $\omega$, $\Omega$), down to a given limit depending on
the star characteristics, e.g.\ 0.5, 1 or 5\,\MEarth. The astrometric
amplitude, $A$, of a \Mstar mass star due to the reflex motion in
presence of a \Mplanet mass planet orbiting around with a semi-major
axis $a$ at a distance $D$ from the Sun is
\begin{equation}
  \label{eq:astrometric-signal}
  A = 0.3 \left(\frac{\Mplanet}{1\,\MEarth}\right) \left(\frac{a}{1\,\AU}\right) 
  \left(\frac{\Mstar}{1\,\Msun}\right)^{-1}
  \left(\frac{D}{10\,\pc}\right)^{-1}  \uas. 
\end{equation}
To detect such a planet, one needs to reach a precision $\sigma =
A/\SNR$ with a typical signal-to-noise ratio $\SNR=6$. If $\sigma_0$
is the precision that NEAT can reach in one single observation that
lasts $t_0$ (e.g.  $\sigma_0 = 0.8\,\uas$ in $t_0 = 1$\,h), when
observing the same source \Nvisits times during \Tvisits each visit
requires
\begin{equation}
  \label{eq:Tvisits}
  \Tvisits = t_0 \left(\frac{A}{\SNR\,\sigma_0}\right)^{-2} (2\Nvisits - m)^{1/2}
\end{equation}
for a given \Nvisits, and with $m = 5+7p$ parameters where $p$ is the
number of planets in the system since there are 5 parameters
characterizing the star astrometric motion and 7 parameters for each
orbit. $\Nvisits \approx 50$ is sufficient to solve for the parameters
of 3 to 5 planets per system, for a 5-yr duration of the mission.

\subsection{Targets}
\label{sec:mission-definition}

\begin{figure}[t]
  \centering
  \hfill
  \parbox[c]{0.4\hsize}{\includegraphics[width=\hsize]{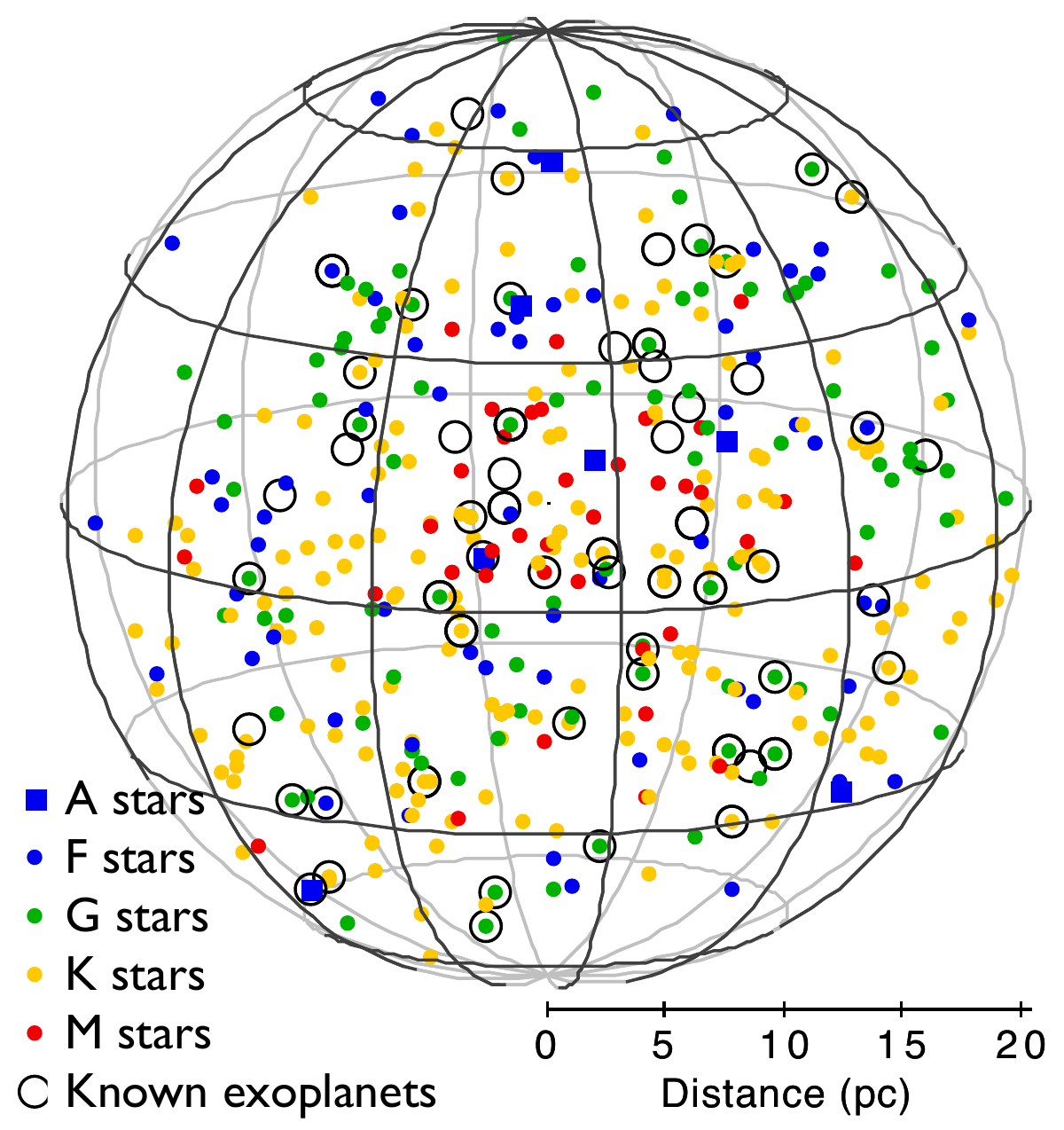}}
  \hfill
  \parbox[c]{0.25\hsize}{\fbox{\includegraphics[width=\hsize]{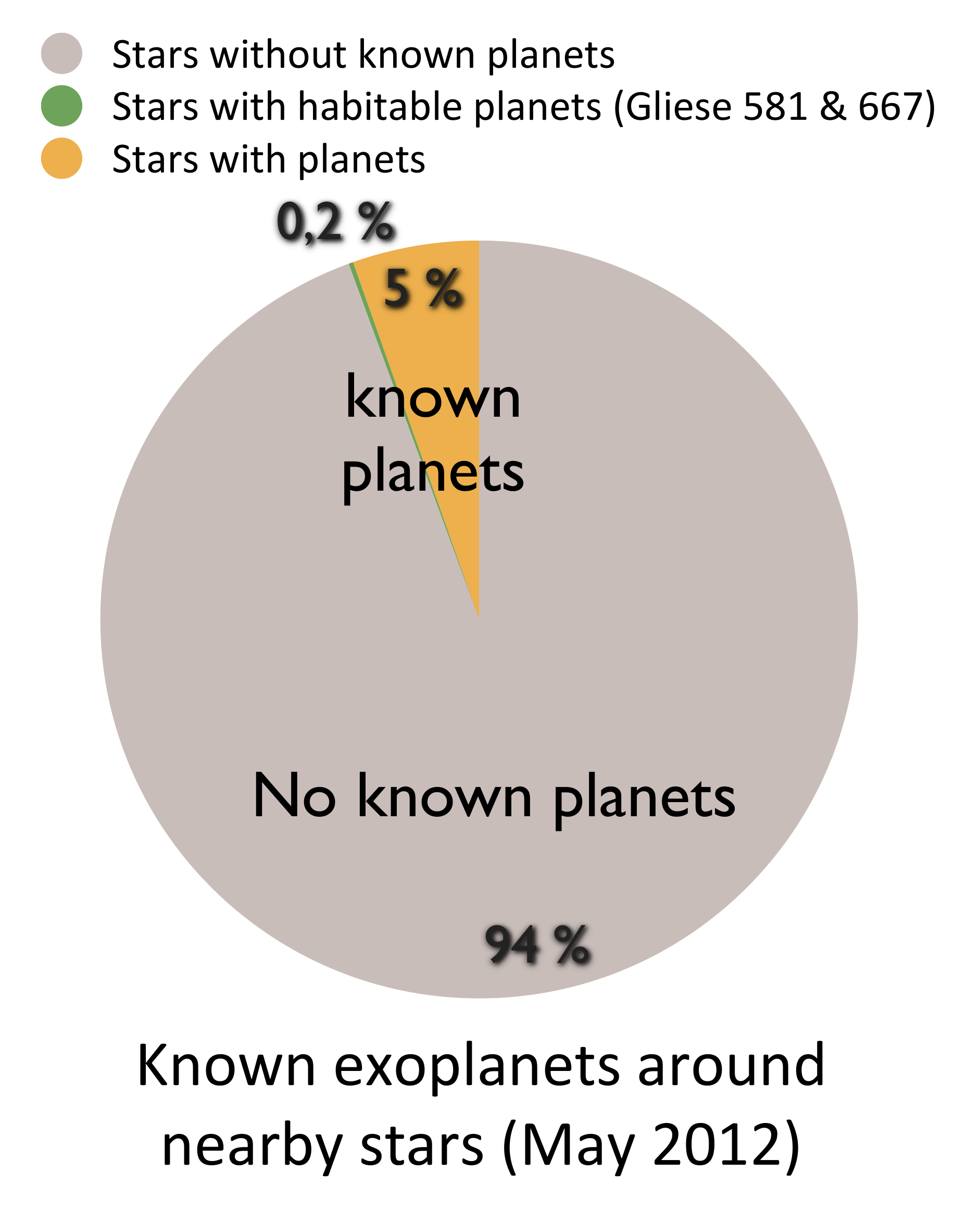}}}
  \hfill ~ \\*[2em]
\caption{Left: representation of the NEAT targets in the 3D sphere of our
    neighborhood ($D$ up to $\approx20\,\pc$). Right: number of these
    stars with known exoplanets. Only 5\% of these stars are known to
    host exoplanets.\bigskip}
  \label{fig:targets-sphere}
\end{figure}

The spatial repartition of stars within 20\,pc is shown in
Fig.~\ref{fig:targets-sphere}. As for the date of May 2012, only 5\%
are known to host exoplanets because of stellar activity problems and
brightness issues. Most of the more than 700 exoplanets known in May
2012 are located further away than the first 20\,pc.

\begin{table*}[t]
  \centering
  \caption{Partial list of possible targets.  Stars are ranked by
decreasing astrometric signal for a planet in its habitable zone
(HZ). This signal $A$(\uas) is calculated for 0.5, 1 and 5\,\MEarth
planets around the 5, 70 and 200 first stars, respectively, assuming
that the planet is located at the inner boundary of the HZ that
secures its detection whenever the planet is in this zone. The
corresponding integration time ($t_{\rm visit}$ in h) and cumulated
times ($t_{\rm tot}$ in h) are calculated for a detection with an
equivalent $\SNR = 6$.}
  \smallskip
  \includegraphics[width=0.95\hsize]{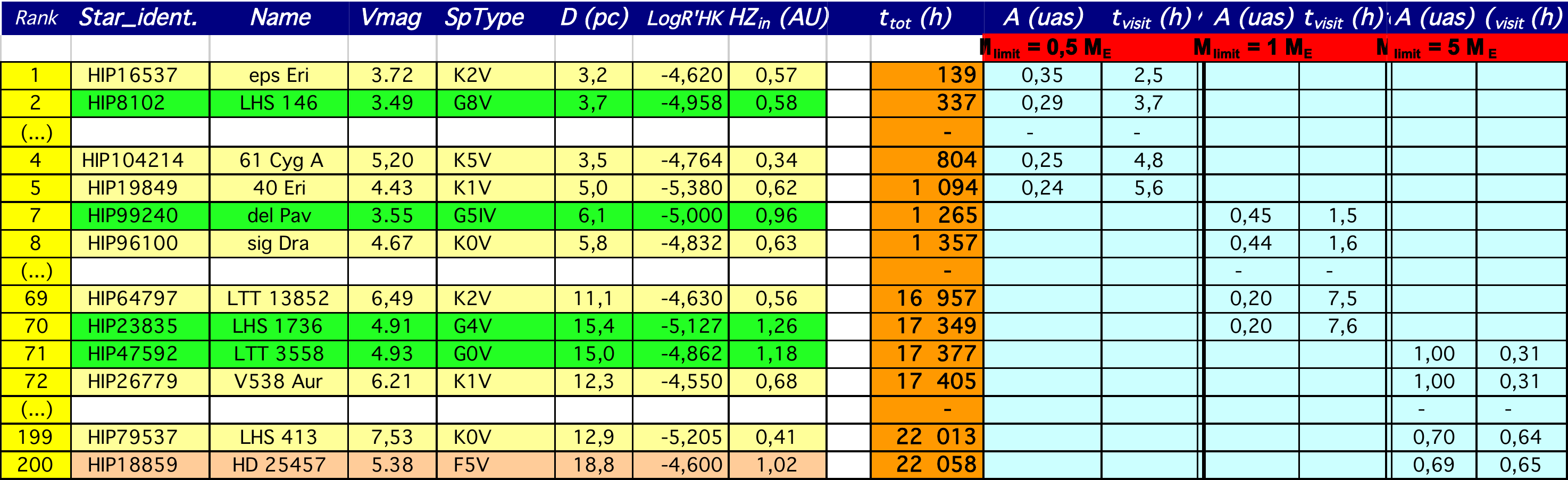}
\label{tab:target-list}
\end{table*}

A possible target list is shown in Table~\ref{tab:target-list} where
we consider the list of the nearest F, G, K stars deduced from the
Hipparcos 2007 catalogue (new data reduction\cite{2007A&A...474..653V}
), disregarding spectroscopic binaries, and stars with an activity
level 5 times greater than that of the Sun because of their
astrometric noise (only 4\% of the sample \cite{2011A&A...528L...9L} )
and for which we compute the astrometric signal for a planet with
given mass in the HZ of the stars
\cite{2010AsBio..10..103K}. Conservatively, we select the inner part
of the HZ in order to be able to detect the planet whatever is its
location in the HZ. The required number of visits and cumulative time
to observe this list of target stars is summarized in
Table~\ref{tab:program}.
\begin{table}[t]
  \centering
  \caption{Left: summary of the main program capabilities and required
    resources. Right: Time and allocated maneuvers for the different
    programs: (1) the Gaia Mission and its Exoplanet Science Potential;
    (2) NEAT follow-up program of Gaia detected  planetary systems;
    (3) observations of young stars; and (4) characterizing planetary
    systems around some of the closest A and M stars.} 
  \label{tab:program}
  \smallskip
    \includegraphics[width=0.4\hsize]{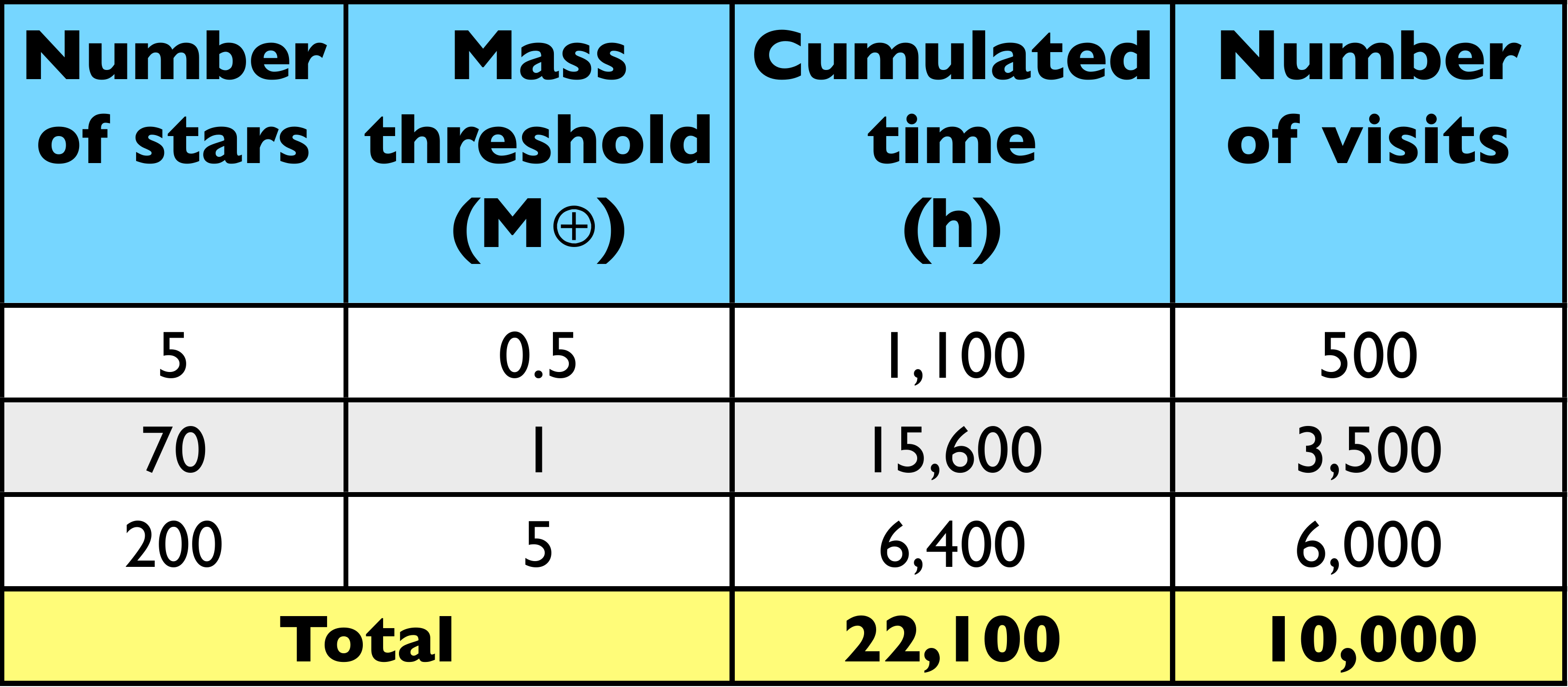}
    \includegraphics[width=0.44\hsize]{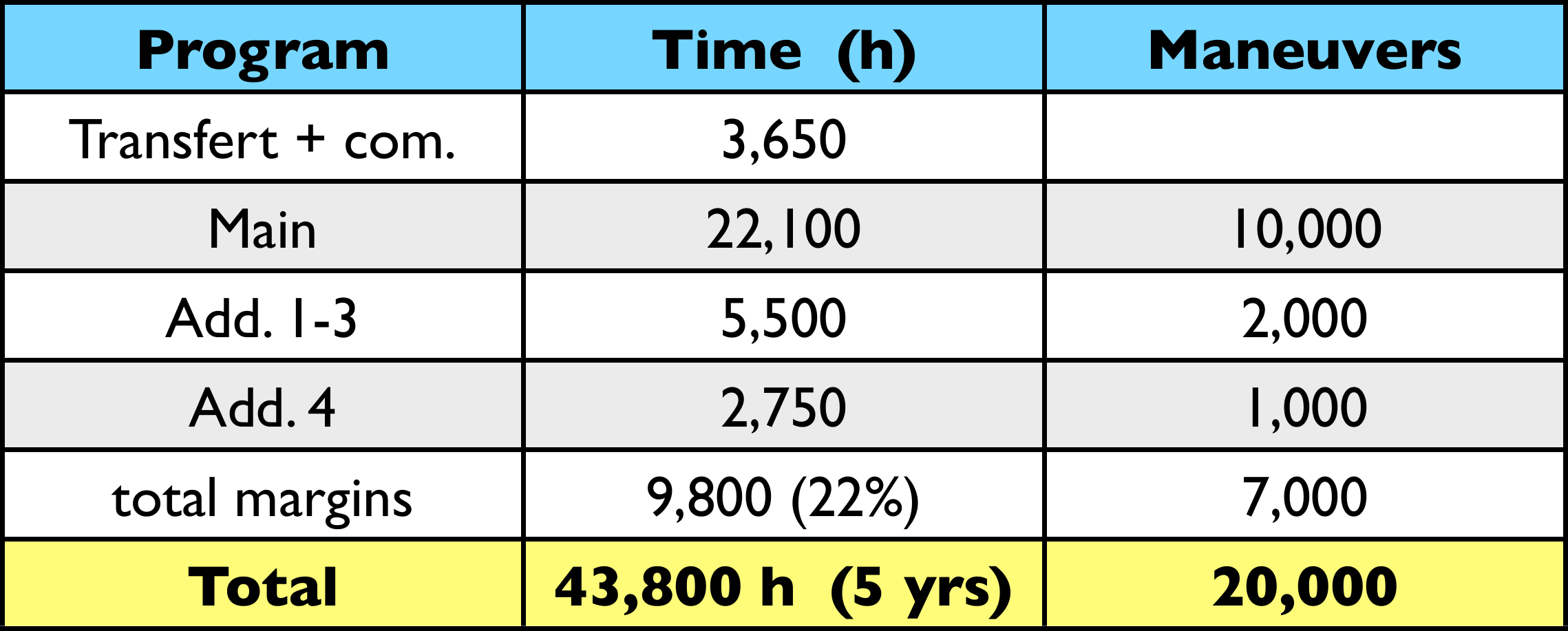}
\end{table}
The list corresponds to an exhaustive search for 1 Earth mass planets
(resp.\ 5 Earth mass planets) around K stars up to 6 pc (resp.\ 12
pc), G stars up to 10 pc (resp.\ 17 pc), and F stars up to 14 pc
(resp.\ 19 pc) in the whole HZ of the star, excluding spectroscopic
binaries and very active stars. 

60\% of the NEAT targets (118) are brighter than $V=6$ and therefore
will not be investigated by Gaia because of its bright limit. So, even
if some of those sources do not harbor Earth-like planets, NEAT will
be contributing to the improvement of our knowledge about the
neighborhood of our Solar System. In that respect, NEAT observations
will not only be complementary to Gaia's ones, but NEAT data will also
form a base to improve Gaia results.

In addition to the survey for the NEAT main science program, we
propose that 30\% of NEAT time is allocated to study some objects of
interest (planets around A stars and M dwarfs, young stars, multiple
systems,... discovered by Gaia and others). The global required amount
of time and number of maneuvers is listed in the right
Table~\ref{tab:program}.

\section{NEAT concept}
\label{sec:neat-concept}

Our goal is to detect the signal corresponding to the reflex motion of
a Sun-like star at 10\,pc due to an Earth-mass planet in its HZ, with
an equivalent final SNR of 6. That astrometric signal is
0.3\,\uas. The required end-of-mission noise floor is 0.05\,\uas, over
100 times lower than Gaia's best precision which is 7\,\uas.

\subsection{Technical challenges}
\label{sec:technical-challenges-1}

Achieving sub-micro-arcsecond astrometric precision, e.g. 0.8\,\uas,
in 1 hour and a noise floor under 0.05\,\uas with a telescope of
diameter $D$ requires mastering all effects that could impact the
determination of the position of the point spread function. The
typical diffraction limited size of an unresolved star is about
$1.2\lambda/D$, which corresponds to 0.16 arcseconds for a 1-m
telescope operating in the visible spectral region.  Even though
differential astrometry of stars within the same field of view softens
somewhat the requirement, this level of accuracy can only be obtained
in an atmosphere-free space environment.

Sub-micro-arcsecond level astrometry requires solutions to four
challenges:
\begin{itemize}
\item \textbf{Photon noise.} Most target stars are bright
  ($R\leq6$\,mag), but the associated reference stars are faint,
  therefore their brightness dominates the photon noise.  Using the
  mean stellar density in the sky, one finds that a field of view
  (FOV) as large as diam $0.6^\circ$ is needed to get several (6 to 8)
  of $R\leq11$\,mag references (see e.g.\ Fig~\ref{fig:ups-and}).
\item \textbf{Beam walk.} A classical three mirror anastigmat (TMA)
  telescope can manage a $0.6^\circ$ diffraction limited
  FOV. However the light coming from different stars, and therefore
  from different directions, will hit the secondary and tertiary
  mirrors on very different physical parts of the mirrors. The mirror
  defects will therefore produce different and prohibitive astrometric
  errors between the images of the stars. Using a single mirror
  telescope solves this problem. To obtain sufficiently high angular
  resolution, a long focal length for this mirror is
  needed, with no intermediate mirrors, a relatively unusual solution
  in modern optical astronomy.
\item \textbf{Stability of the focal plane.} Proper Nyquist sampling
  with typical detector pixels of the order of 10\,\microns requires a
  focal plane at a focal length of 40\,m. Such a focal plane covering
  a FOV of $0.6^\circ$ diameter would yield a costly detector mosaic
  with $40,000\times40,000\approx10^9$ pixels. Sub-microarcesc
  astrometry over a $0.6^\circ$-diameter FOV requires the geometry of
  the focal plane to be stable to
  $\approx1:2\times10^{-10}$. Instead of building a gigapixel focal plane
  with unprecedented stability we plan to use 9 small $512\times512$
  CCDs (right part of Fig.~\ref{fig:payload-concept}) and a laser metrology
  system to measure the position of every pixel to the required
  precision, once every 10 to 30\,s. We do not rely on their
  positioning and stability, but we measure it accurately and
  frequently with a laser metrology based
  on dynamic interference fringes.
\item \textbf{Quantum efficiency (QE) variations.} At the level of
  accuracy required by this mission, the measurement of the inter- and
  intra-pixel QE variations is mandatory. Thanks to a the dynamic
  fringes of the calibration system, each pixel response can be
  characterized\cite{2011RSPSA.467.3550Z,2011SPIE.8151E..28N} with six
  parameters such that the systematic errors are kept below $10^{-6}$.
\end{itemize}

\subsection{Instrumental concept}
\label{sec:instrumental-concept}

\begin{figure}[t]
  \centering
  \includegraphics[width=0.68\hsize]{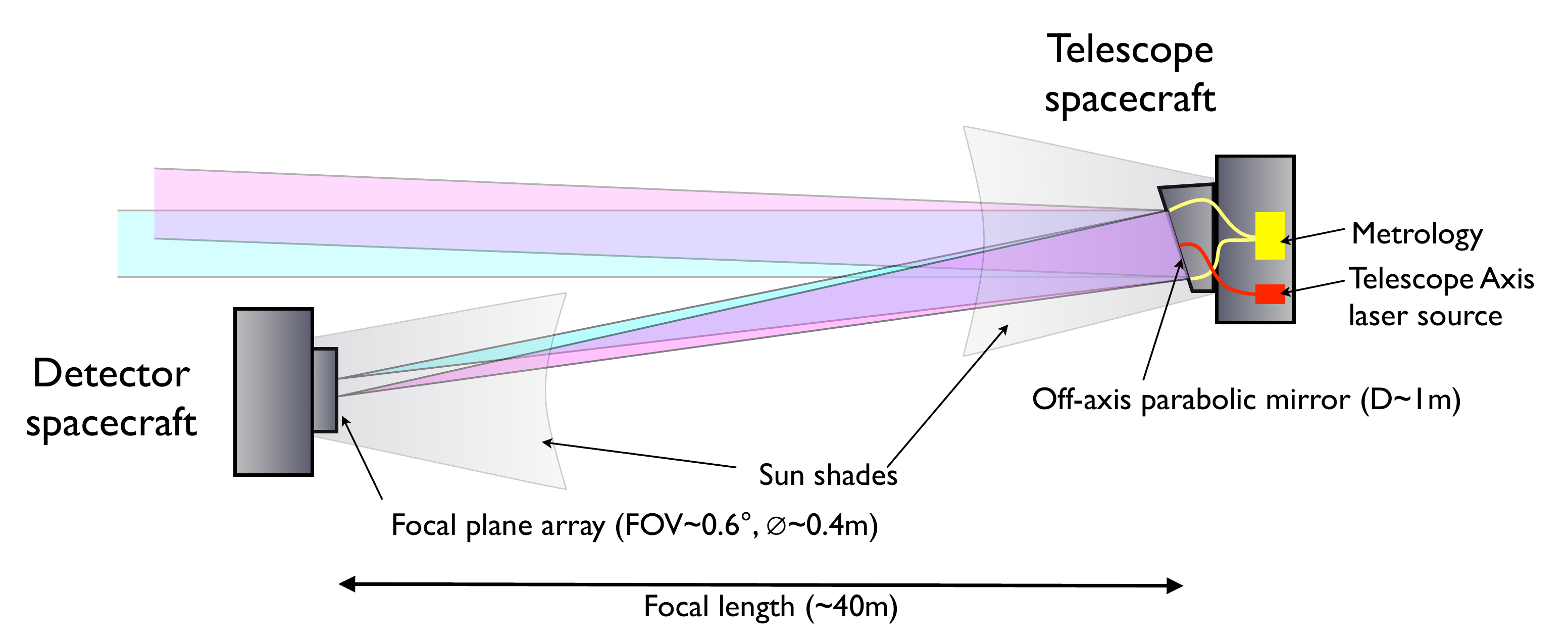}
  \hfill
  \includegraphics[width=0.3\hsize]{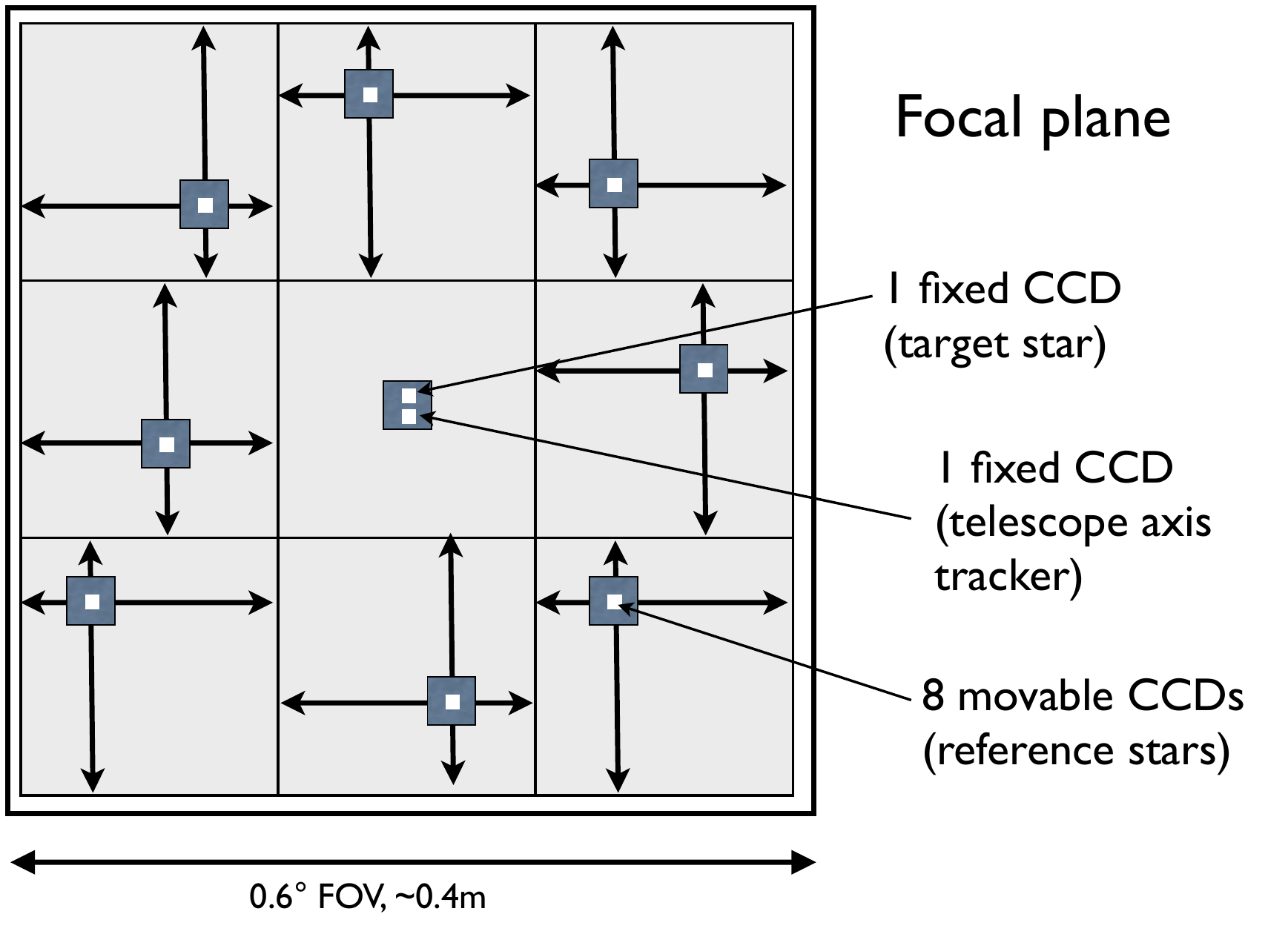}  
  \caption{Proposed concept for a very high precision astrometry
    mission. Left: the instrument consists in two separated modules,
    the first one carrying the primary mirror (upper right) and the
    second one the detector plane (bottom left). Right: schematic
    layout of the focal plane. The field of view is divided in
    $3\times3$ sub-fields. Outer subfields have visible arrays
    which can be moved in $X$ and $Y$ directions to image the reference
    stars. The central field has two fixed arrays, one for the target
    star and one for the telescope axis tracker.}
  \label{fig:payload-concept}
\end{figure}

The proposed mission is based on a concept that results from the
experience gained in working with many astrometry concepts (SIM,
SIM-Lite, corono-astrometry\cite{2010SPIE.7731E..68G}).  The concept
is sketched in left part of Fig.~\ref{fig:payload-concept} and
consists of a primary mirror ---an off-axis parabolic 1-m mirror--- a
focal plane 40 m away, and metrology calibration sources. The large
distance between the primary optical surface and the focal plane can
be implemented as two spacecraft flying in formation, or a long
deployed boom. The focal plane with the detectors having a field of
view of $0.6^\circ$ is shown in the right part
Fig.~\ref{fig:payload-concept}. It has a geometrical extent of
$0.4\meters\times0.4\meters$. The focal plane is composed of eight
$512\times512$ visible CCDs located each one on an $XY$ translation
stage while the central two CCDs are fixed in position.  The CCD
pixels are 10\microns in size.

\begin{figure*}[t]
  \centering
  \includegraphics[trim= 0em 8em 6em 6em, clip,width=0.8\hsize]{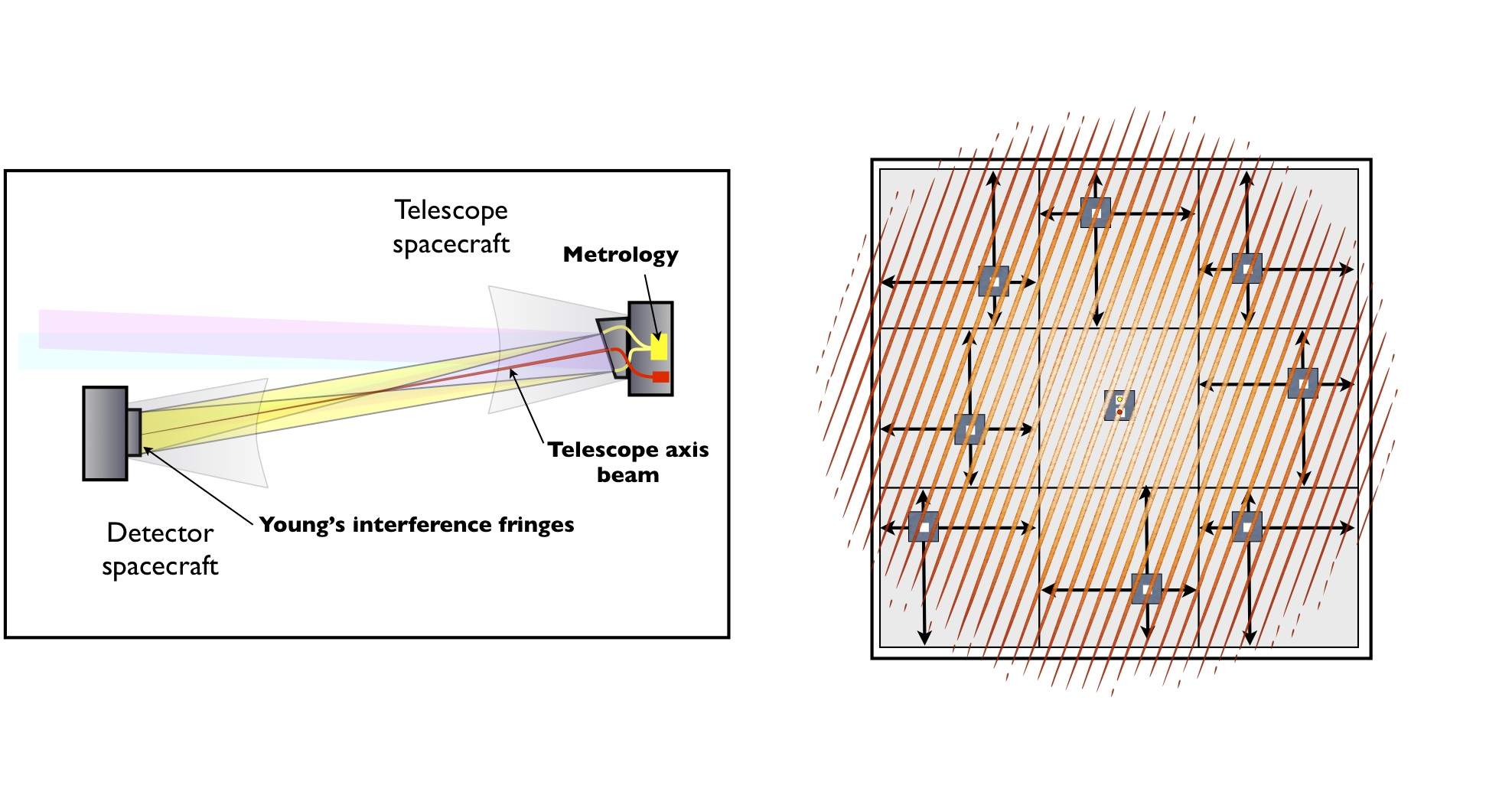}
  \caption{Principle of the metrology and the axis tracker. Left
    panel: the metrology laser light (in yellow) is launched from
    fibers located at the edge of the mirrors. Right panel: the laser
    beams interfere over the detector plane. Only the fringes
    corresponding to a pair of fibers are represented on this figure
    and they are not to scale, since the fringe spacing is equal to
    the PSF width. The axis tracker (sketched in red on the left
    panel) is a laser beam launched in the center of the mirror that
    is monitored in the lower central CCD.}
  \label{fig:metrology-concept}
\end{figure*}

The principle of the measurement is to point the spacecraft so that
the target star, which is usually brighter ($R\leq6$) than the
reference stars ($R\leq11$), is located on the axis of the telescope
and at the center of the central CCD. Then the 8 other CCDs are moved
to center each of the reference stars on one of them. To measure the
distance between the stars, we use a metrology calibration system that
is launched from the telescope spacecraft and that feeds several
optical fibers (4 or more) located at the edge of the mirror. The
fibers illuminate the focal plane and form Young's fringes detected
simultaneously by all CCDs (Fig.~\ref{fig:metrology-concept}). The
fringes have their optical wavelengths modulated by acoustic optical
modulators that are accurately shifted by 10\,Hz, from one fiber to
the other so that fringes move over the CCDs\cite{2011SPIE.8151E..28N,
  2012SPIE.8442.....N} or by phase
modulator\cite{2012SPIE.8445.....C}.  These fringes allow us to solve
for the $XYZ$ position of each CCD. An additional benefit from the
dynamic fringes on the CCDs is to measure the QE of the pixels (inter-
and intra-pixel dependence). The CCDs are read at 50\,Hz providing
many frames that will yield high accuracy.

With the proposed concept, it is possible to achieve all the main technical
requirements:
\begin{itemize}
\item \textbf{Focal plane stability.} Instead of maintaining a focal
  plane geometry stable at the 0.1\,nm level for a 5-yr duration,
  which is impossible, we get advantage of a metrology for every pixel at the
  sub-nanometer level which is used every minute.
\item \textbf{Reference frame.}  By measuring the fringes at the
  sub-nanometer level and by using the information from all the pixels of
  each CCD, it is possible to solve for the
  position of all reference stars compared to the central target with
  an accuracy of $0.8\,\uas/\sqrt{h}$. The
  field of view of $0.6^{\circ}$ allows us to have 6 to 8 reference stars
  brighter than $V=11$ in most fields.
\item \textbf{Photon noise.} The field of $0.6^{\circ}$ provides about 6 to
  8 stars of magnitude brighter than $R=11$. The number of photons
  received by one 11-mag star on the system is $\approx
  4.1\times10^9\,\mbox{ph/hr}$. Since the FWHM of diffraction-limited
  stars is $1.2\lambda/D=0.16$ arcsecond, the photon noise limit in
  1\,h of integration due to a set of 6 reference stars is
  $(\lambda/2D)/\sqrt{6N}\approx 0.5\,\uas$.  With more than
  $50\times2$ measurements of a few hours spread over 5\,yrs, the
  equivalent precision is 0.05\,\uas in RA and Dec, corresponding to
  the detection of the 0.30\,\uas signal with a SNR $\approx 6$.
\item \textbf{Large-scale calibration.} The detector plane does not
  have to be fully covered by pixels, since the positions of the
  reference stars are known from available catalogues (10-20\,mas for
  Tycho 2, and about tens of \uas for Gaia). For the target stars
  ($R\leq6$), we use the Hipparcos catalog (few mas accuracy). This
  corresponds to $<1/10^{\rm th}$ of the PSF or the fringe width. The
  number of fringes between the target star and the reference stars is
  then known, only the positions of the star centroids relative to the
  interferometric fringes have to be measured accurately.
\end{itemize}
The use of 10 small CCDs drastically reduces the cost of what would
otherwise be a giga-pixel focal plane and also helps to control
systematics. With such a concept, the mission performance
would be similar to, and even more favorable for exoplanets, than what
was proposed for SIM-Lite with 5 years of operation, but at the price
of giving up all-sky astrometry and the corresponding science
objectives.

\subsection{Performance assessment and error budget}
\label{sec:perf-assessm-error}

NEAT will not be capable of measuring the absolute separation between
the target and the set of reference stars to 0.8\,\uas.  NEAT
objectives is to measure the change in the \emph{relative} position of
those stars between successive observations spread over the mission
life, with an error of 0.8\,\uas for each one hour visit. Achieving
our target precision relies on not only the metrology stability, but
also on the precise knowledge of the positions of the multiple
reference stars used since the expected motions of the references
cannot be considered as fixed (see discussion in
Sect.~\ref{sec:astro-challenges}).  Our comprehensive error budget
takes into account all sources of error, including instrumental
effects, photon noise and astrophysical errors in the reference star
positions.

\begin{figure*}[t]
  \centering
  \includegraphics[width=\hsize]{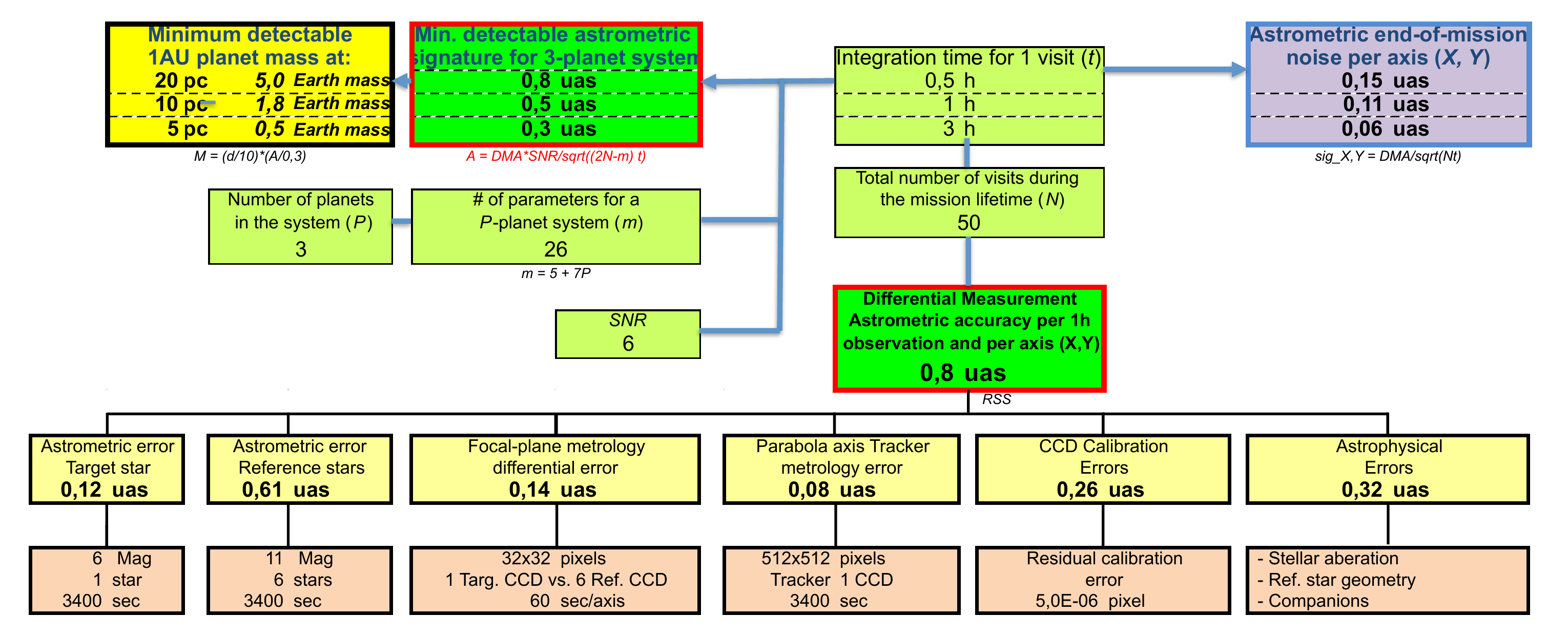}
  \caption{Top-level error budget for NEAT that shows the major
    contributor to the overall budget. It also shows how the
    0.8\,\uas/$\sqrt{h}$ accuracy enables the detection of 0.3\,\uas
    (resp.\ 0.5\,\uas and 0.8\,\uas) signatures with a signal to noise
    of 6 after 50 visits of 3\,h (resp.\ 1\,h and 0.5\,h) of
    observation that allows a 0.5\,\MEarth planet at 5\,pc to be
    detected (resp.\ 1.8\,\MEarth at 10\,pc and 5\,\MEarth at 20\,pc).}
  \label{fig:error-budget}
\end{figure*}

The six major errors terms are captured in the simplified version of
the error budget shown in Fig.~\ref{fig:error-budget}. The biggest
term is the brightness dependent error for the set of reference stars.
Static figure errors of the primary mirror will produce centroid
offsets that are mostly common-mode across the entire field of
view. Differential centroid offsets are significantly smaller than the
field-dependent coma and are in fact negligible.  Similarly, changes
in the primary mirror surface error, e.g.\ due to thermal dilatation,
meteorite impacts,... produce mostly common-mode centroid shifts and
negligible differential centroid offsets. On the other hand,
displacement and changes in the shape of the PSF would couple with the
CCD response if the CCD response is not properly calibrated. This is
continuously done by the metrology fringes.

\subsection{Design of the payload subsystems}
\label{sec:design-payload-subsystems}

With all these constraints in mind, we have designed the NEAT
subsystems which will permit to achieve the main requirements.

\begin{figure*}[t]
  \centering
  \includegraphics[trim=20mm 20mm 20mm 0mm,clip, width=0.9\hsize]{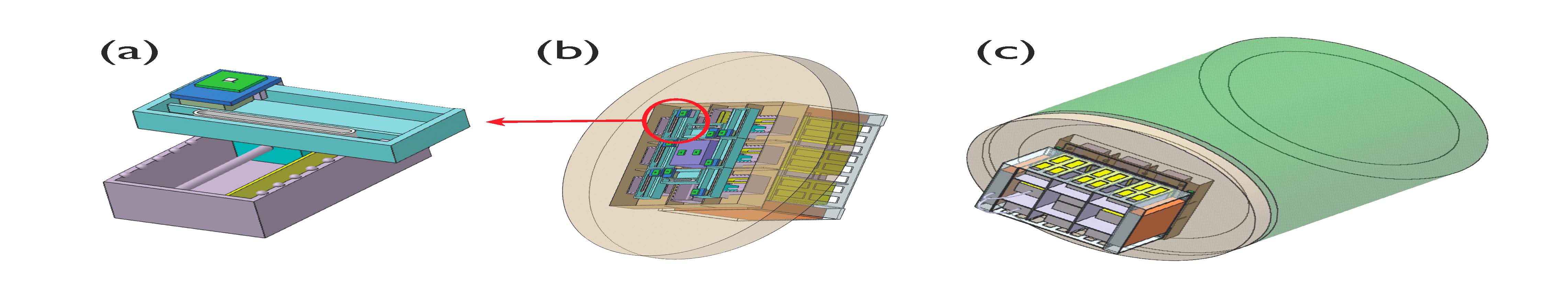} 
 \caption{Views of the focal plane assembly. (a) Magnified
    view of one of the 8 $XY$ translation stages of the focal plane. In
    yellow, the $512\times512$ CCD and its support. In blue and green the two
    translation stages. (b) The front part of the focal plane
    with its 8 movable CCDs and two fixed CCDs at the center. (c) The electronics racks.}
  \label{fig:focal-plane-drawing}
\end{figure*}

\textbf{Focal plane assembly}. A proposition for implementation of the
focal plane is shown in Fig.~\ref{fig:focal-plane-drawing}.
The detector is foreseen to be a CCD fabricated by the E2V
technologies company in UK. The target star, the reference star and
the telescope axis tracker will all use the same CCD that includes the
capability to read windowed images, typically $10\times10$ to
$30\times30$ pixels. The 8 $XY$ tables consist of two linear tables
mounted on top of each other. Each table uses a piezo-reptation
motor\footnote{such reptile motors have been qualified by the Swiss
  firm RUAG for the LISA GPRM experiment.}, a linear ball bearing
system and an optical incremental encoder. These motors fulfill
several requirements of simplicity: they are self-locked when they are
not powered; they can be used both for large displacements by stepping
up to $100\,\mbox{mm}\times100\,\mbox{mm}$ and elementary analog
motion down to 50\,nm.  Since 8 tables are used in parallel in the
focal plane, the loss of one table is not a single point failure. An
alternative implementation could be to drive the $XY$ tables with ball
screws and rotary motors. The limited resolution of such a motor stage
(about 5\,\microns) could be supplemented by a second high-resolution
piezo $XY$ table\footnote{such as the Cedrat XY25XS}, mounted on top
of the first $XY$ table.  The main structure of the focal $XY$ tables
consists of a large lightweight aluminum cylinder which is thermally
controlled and in which pockets are machined for the fixation of the
$XY$ tables.

\textbf{Telescope}. The primary mirror is an off-axis paraboloid, with
a 1\,m diameter clear aperture, an off-axis distance of 1\,m and a
focal length of 40\,m. It would be manufactured in either Zerodur or
ULE, 70\% light-weighted and weight about 60\,kg. The surface quality
should be better than $\lambda/4$ peak-to-valley and would be coated
with protected aluminum. A 50\,mm hole at the center of the mirror
accommodates the beam launcher for the telescope axis tracker.  The
three bipods on the back of the mirror support the mirror with minimum
deflection. The bipods interface with the tip-tilt stage made of 3
preloaded piezo stacks on parallel flexures that provide the $\pm6$
arcsecond amplitude for two-axis articulations. The entire primary
mirror assembly interface to the telescope payload plate is a 34\,kg
hogged-out aluminum plate. This plate also hosts the metrology source,
the telescope drive electronics, the telescope baffle and the
interface to the spacecraft.

\textbf{Telescope axis tracker}. It is used to estimate the location
of the primary mirror axis with respect to the focal plane in order to
monitor it and then correct for the telescope field dependent
errors. The sensor is the second fixed CCD located in the
focal-plane. The launcher would consist of either an achromatic
doublet or an aspherical singlet lens, embedded in the primary mirror
substrate at its center and a single-mode fiber-coupled laser diode.

\textbf{Laser metrology}. The focal-plane metrology system consists of
the metro\-logy source similar to the one developed for SIM
\cite{2010SPIE.7734E..71E}, the metrology fiber launchers and the
focal plane detectors (CCDs) which alternatively measure the stellar
signal (57\,s observations) and the metrology (1\,s per axis every
minute). The metrology fiber launchers consist of nominally four
optical fibers attached to the primary mirror substrate. Three of them
are located around the edge of the mirror. They are used by pairs in
order to conduct three non-redundant measurements of the relative
location of the CCDs. The fourth fiber is located inside the clear
aperture, and is used in combination with each of the three other
fibers to produce three additional measurements during focal plane
calibration and calibrate the distance mirror-focal plane.

\textbf{Pointing servo systems}. The pointing of the telescope from
one target to the next one is accomplished by the two spacecraft in
formation flying. The target stars will be typically separated by
$10^{\circ}$. Re-pointing of the telescope will require rotation of the two
spacecraft by several degrees using reaction wheels and translation of
the telescope spacecraft by several meters using hydrazine
propulsion. Fine positioning of the focal plane relative to the mirror
is done by cold gas propulsion system. At the end of the maneuver,
the telescope spacecraft will be oriented to better than 3 arcseconds
from the target star line of sight using star trackers. The focal
plane spacecraft will be positioned to better than 2\,mm from the
primary mirror focus. At that point, the spacecraft will maintain their
relative position to better $\pm$2\,mm in shear and in separation for
the duration of the observation. The separation does not require a
servo-loop of the payload, because its effect is only a degradation
in performance (when FWHM increases, final precision decreases in same
proportion) and is managed in the error budget.

During the observation, the instrument uses a tip-tilt stage behind
the primary mirror to center the target star on the $32\times32$ pixel
sub-window on the target star CCD.  Once in the $32\times32$ pixel
sub-frame mode, the target star CCD is read at 500\,Hz, and feedback
control between the CCD and the tip-tilt stage can be used to keep the
star centered on the detector to better than 5 milli-arc-second RMS
(0.1 pixel RMS) for the duration of the observation. This is the only
active feedback loop in the instrument system working at 50\,Hz; the
other degrees of freedom (focal plane tip, tilt, clocking and
focal-plane-to-mirror separation) are monitored but not corrected for
in real-time.  Prior to acquisition, the reference star CCDs will be
pre-positioned to the expected location of the reference stars using
the translation stages. The $XY$ translation stage fine motion of the
reference star CCDs at a 0.2\,\microns precision enables centering of
the reference stars on the detectors to better than a tenth of a
pixel. Once the reference stars are acquired, the translations stages
are locked for the duration of the observation.

\subsection{Mission requirements}
\label{sec:spacecrafts}

The objectives of the NEAT mission require to perform acquisitions
over a large number of targets during the mission timeline, associated
to a 40\,m focal length telescope satellite. The preliminary assessment of the
NEAT mission requirements allows to identify the following main
spacecraft design drivers. 

\textbf{Launch configuration and mission orbit.} The L2 orbit is the
preferred orbit, as it allows best formation flying performance and is
particularly smooth in terms of environment. The Soyuz launch,
proposed as a reference for medium class missions, offers satisfying
performance both in terms of mass and volume.

\textbf{Formation Flying and 40\,m focal length.}  The mission relies
on a 40\,m focal length telescope, for which the preferred solution is
to use two satellites in formation flying. The performance to be
provided by the two satellites in order to initialize the payload
metrology systems are of the order of magnitude of $\pm2$\,mm in
relative motion, and of 3 arcseconds in relative pointing, which are
typically compatible with Formation Flying Units and gyroless
AOCS\footnote{Attitude and Orbit Control System} architecture. At L2,
the solar pressure is the main disturbance for formation flying
control. As a result, surface-to-mass ratio (S/M ratio) is the main
satellite drift contributor and should be as close as possible for the
two satellites. Although satellite design can cope with these
requirements, the S/M ratio of the satellites will evolve during the
mission because of fuel losses and sun angle. However, the
preliminary mission assessment tends to demonstrate that the S/M
difference between the two satellites can be reduced down to 20-30\%,
which is deemed compatible with mission formation flying requirements.

\textbf{Number of acquisitions and Mission $\Delta V$.}  The mission
aims at a complete survey of a large number of targets and the
maximization of the number of acquisitions will be a main objective of
the next mission phases. The mission objectives require a threshold of
20,000 acquisitions (see Sect.~\ref{sec:spacecraft-design} for
details).  In addition, the time allocation for these reconfiguration
maneuvers is quite limited, in order to free more than 85\% of mission
duration for observations.  As a result, the mission is characterized
by a large $\Delta V$ (550 to 880\,m/s) dedicated to reconfigurations,
plus allocations for fine relative motion initialization and control
using the $\mu$-propulsion system. This large number of reconfigurations
is also driving the number of thruster firing, which are
qualified to typical numbers of up to 5,000 to 50,000 with cycling as
required for NEAT.  

\textbf{Baffles and Parasitic Light.}  The mission
performance relies on the ability of the focal plane to receive only
star flux reflected by the telescope satellite. A first requirement is
to implement baffles on the two satellites, coupled by a diaphragm on
the focal plane.  In addition, all parasitic light coming from
telescope satellite reflections should be avoided, thus requiring all
bus elements to be shielded by a black cover.  

Following this
preliminary satellite requirement analysis, a first simple and robust
mission concept has been identified.

\subsection{Preliminary Spacecraft Design}
\label{sec:spacecraft-design}

\begin{figure}[t]
  \centering
  \includegraphics[width=0.9\hsize]{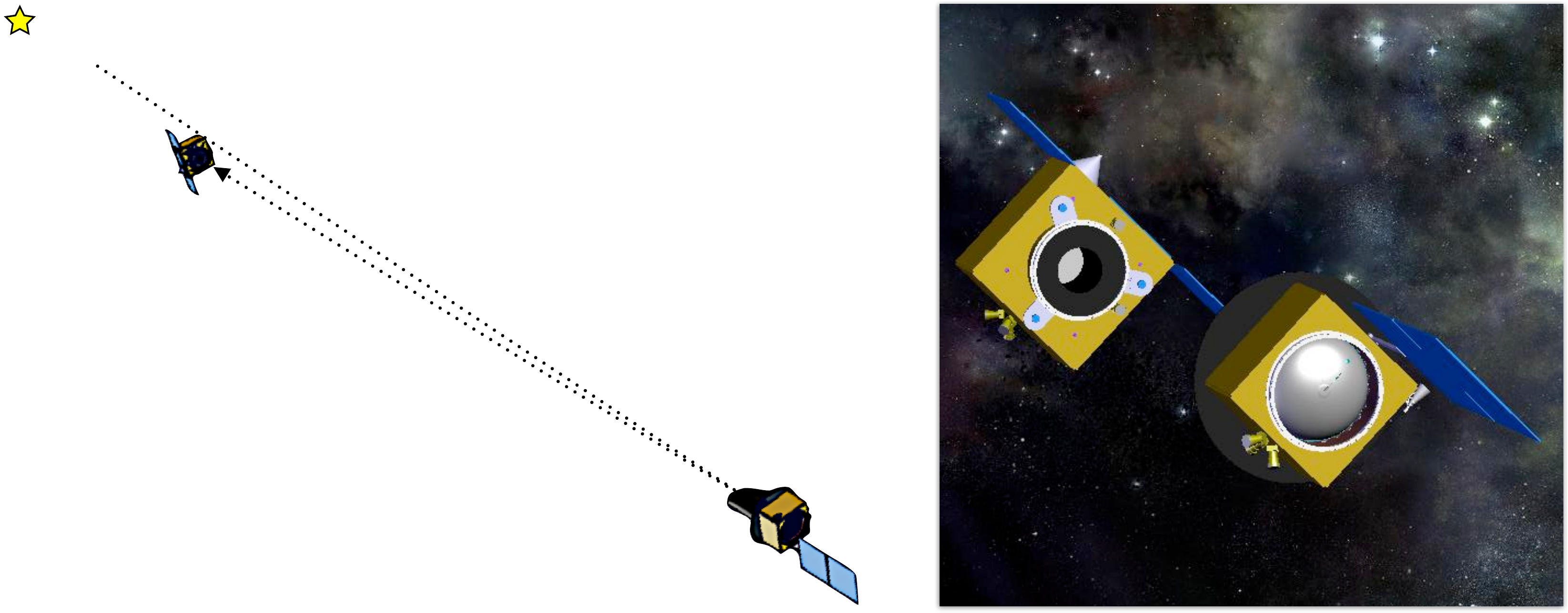}
  \caption{Left: NEAT spacecraft in operation with the two satellites
    separated by 40\,m. Right: closer external view of the two satellites.}
  \label{fig:spacecraft}
\end{figure}

The preliminary NEAT mission assessment allowed to identify a safe and
robust mission architecture (Fig.~\ref{fig:spacecraft}), relying on
high technology-readiness-level (TRL) technologies, and leaving safe
margins and mission growth potential that demonstrates the mission
feasibility within the medium class mission cost cap.

\begin{figure}[t]
  \centering
  \begin{tabular}{ccc}
    \includegraphics[width=0.2\hsize]{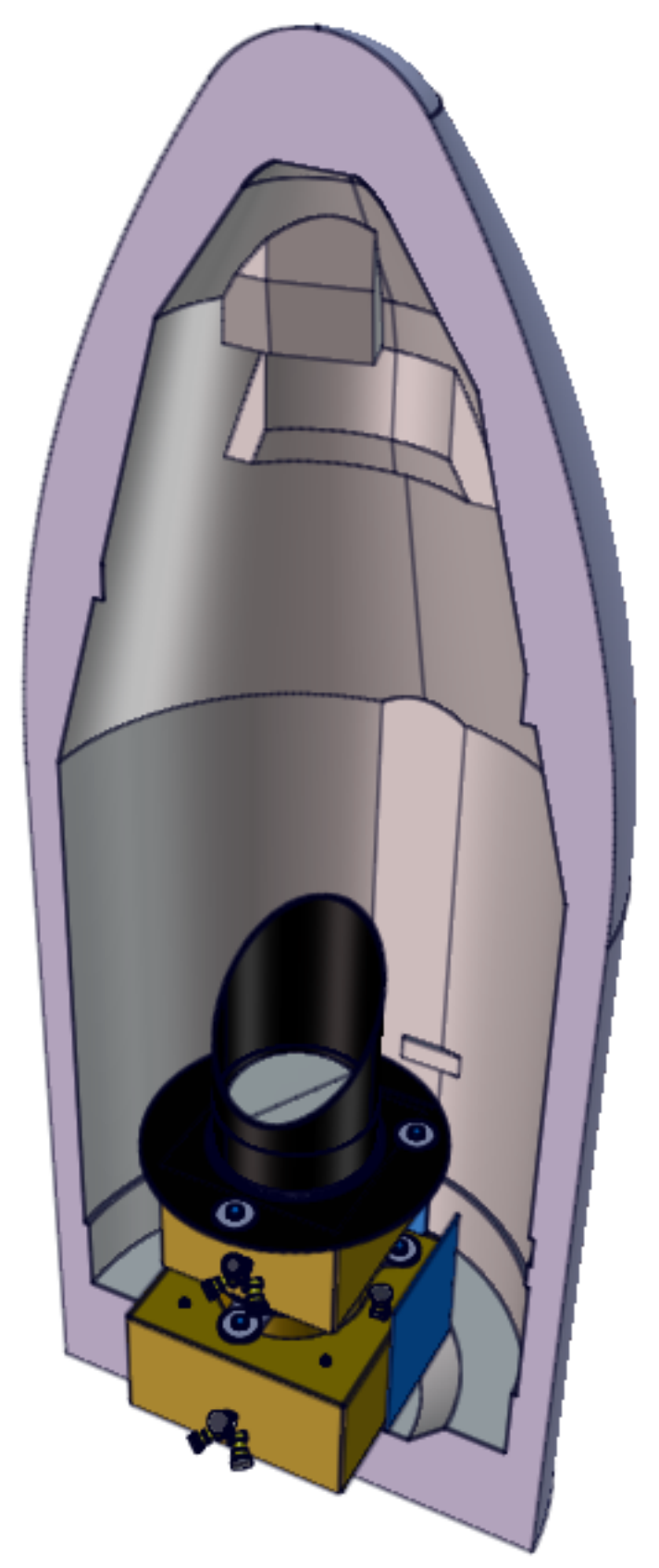}&
    \hspace*{2cm} &
    \includegraphics[width=0.28\hsize]{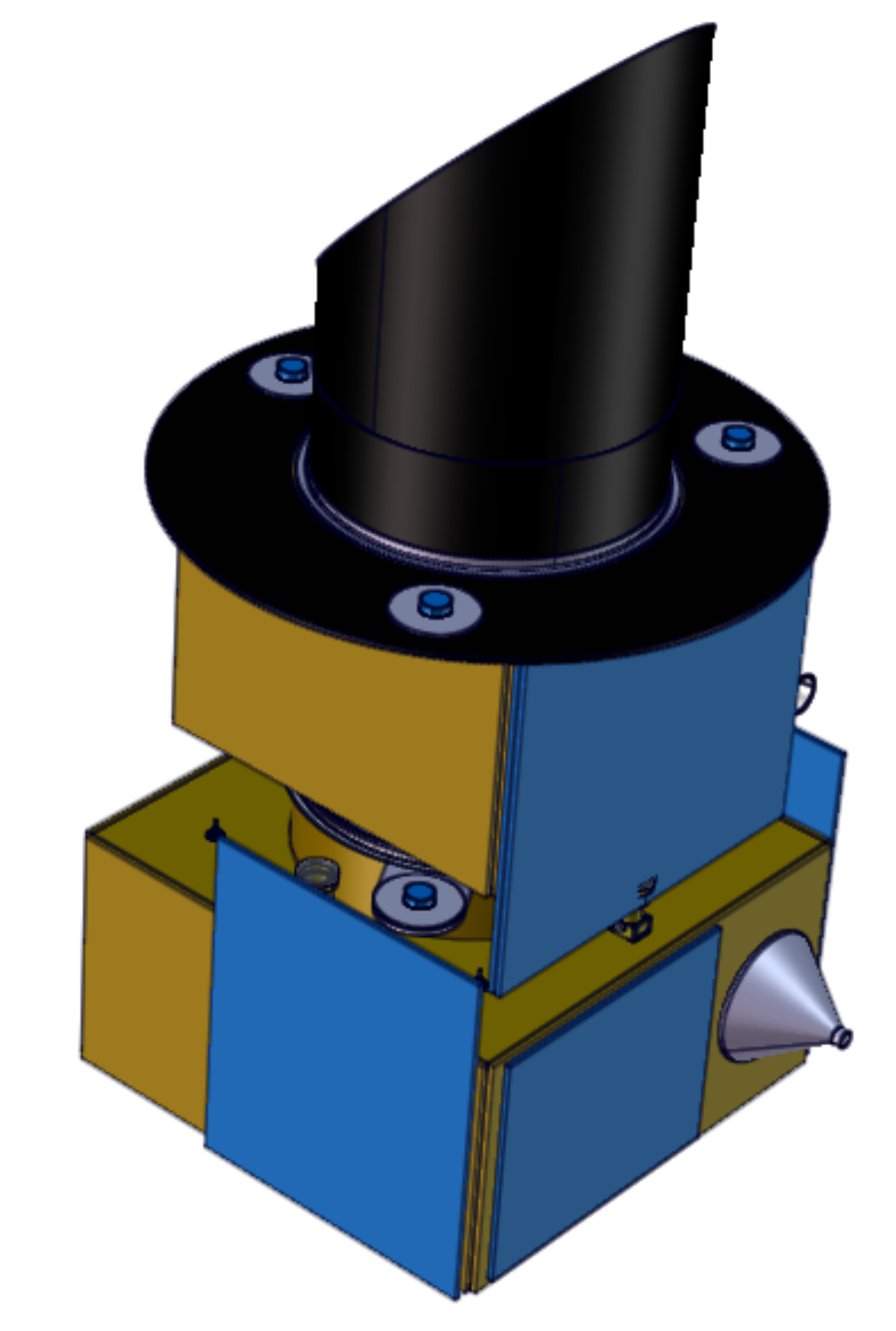}\\
  \end{tabular}
  \caption{NEAT stowed configuration. Left: in the Soyuz
    rocket. Right: back view.}
  \label{fig:stowed}
\end{figure}

\textbf{System Functional Description.}  The proposed mission
architecture relies on the use of two satellites in formation
flying. The two satellites are launched in a stacked configuration
using a Soyuz ST launcher (Fig.~\ref{fig:stowed}), and are deployed
after launch in order to individually cruise to their operational
Lissajous orbit.  Acquisition sequences will alternate with
reconfigurations, during which the Telescope Satellite will use its
large hydrazine propulsion system to move around the Focal Plane
Satellite and to point at any specified star. At the approach of the
correct configuration, the Focal Plane Satellite will use a cold gas
$\mu$-propulsion system for fine relative motion acquisition.  The
Focal Plane Satellite will be considered as the chief satellite
regarding command and control, communications and payload
handling. Communications with the L2 ground station would typically
happen on a daily basis through the Focal Plane Satellite, with data
relay for TC/TM\footnote{Telecommand / Telemetry} from the Telescope
Satellite using the FFRF\footnote{Formation Flying Radio Frequency}
units. This satellite will however be equipped with a similar
communication subsystem, in order to support cruise and orbit
acquisition, and to provide a secondary backup link.

\textbf{Formation Flying Architecture.}  The formation flying will
have to ensure anti-collision and safeguarding of the flight
configuration, based on the successful PRISMA flight heritage. In
addition, the spacecraft will typically perform 12 to 20 daily
reconfigurations of less than $10^{\circ}$ of the system line of sight
corresponding to 7\,m of translation of one satellite compared to the
other perpendicular to the line of sight.  During these
configurations, the telescope satellite will perform translations
---supported by the FF-RF Units--- using its large hydrazine tanks
(250\,kg) for a $\Delta V \approx 605$\,m/s. When the two satellites
will approach the required configuration, the telescope satellite will
freeze, and the focal plane satellite will perform fine relative
pointing control using micro-propulsion system. As a result, the
micro-propulsion will have to compensate for hydrazine control
inaccuracies, which will require large nitrogen gas tanks (92\,kg for
$\Delta V \approx 75$\,m/s). Finally, 28\,kg of hydrazine carried by
the FP satellite allows $\Delta V \approx 55$\,m/s for station keeping
and other operations.

\textbf{Satellite Design Description.} The design of the two
satellites is based on a 1194\,mm central tube architecture, which
will allow a low structural index for the stacked configuration and
provides accommodation for payloads and large hydrazine tanks.  Strong
heritage does exist on the two satellites avionics and AOCS. In
addition, they both require similar function which would allow to
introduce synergies between the two satellites for design,
procurement, assembly, integration and tests. The proposed AOCS
configuration is a gyroless architecture relying on reaction wheels
and high-performance star trackers (Hydra Sodern), which is compatible
with a 3\,arcsec pointing accuracy (see end of
Sect.~\ref{sec:design-payload-subsystems} for payload control). The
satellites communication subsystems use X-Band active pointing
antenna, supported by large gain antenna for low Earth orbit
positioning and cruise, coupled with a 50\,W RF Transmitter. The
active pointing medium gain antenna allows simultaneous data
acquisition and downlink.  A reference solution for the satellite
on-board computer could rely on the Herschel-Planck avionics.

The two satellites would have custom mechanical-thermal-propulsion
architectures. The telescope satellite features a dry mass of 724\,kg
and the focal plane satellite a dry mass of 656\,kg. The focal plane
satellite carries the stacked configuration. The payload (focal plane
+ baffle) are assembled inside a 1194\,mm central tube, which
will also ensure the stacked configuration structural stiffness. The
spacecraft bus, and large cold gas tanks, will be assembled on a
structural box carried by the central tube. The proposed architecture
uses a large hydrazine tank inside the 1194\,mm central tube which
offers a capacity of up to 600\,kg hydrazine, thus allowing both a low
filling ratio and a large mission growth potential. The payload module
---with the payload mirror, rotating mechanisms and baffle--- is then
assembled on the central tube.

\textbf{Proposed Procurement Approach.}  The NEAT mission is
particularly adapted to offer a modular spacecraft approach, with simple
interfaces between payload and spacecraft bus elements.  For both
satellites, the payload module is clearly identified and assembled
inside the structural 1194\,mm central tube. In addition, a large
number of satellite building blocks can be common to the two
satellites, in order to ease mission procurement and tests. This
configuration is particularly compatible with the ESA procurement
scheme.  The payload is made of 3 subsystems: primary mirror and its
dynamic support, the focal plane with its detectors and the
metrology. 

\begin{figure}[t]
  \centering
  \includegraphics[width=0.45\hsize]{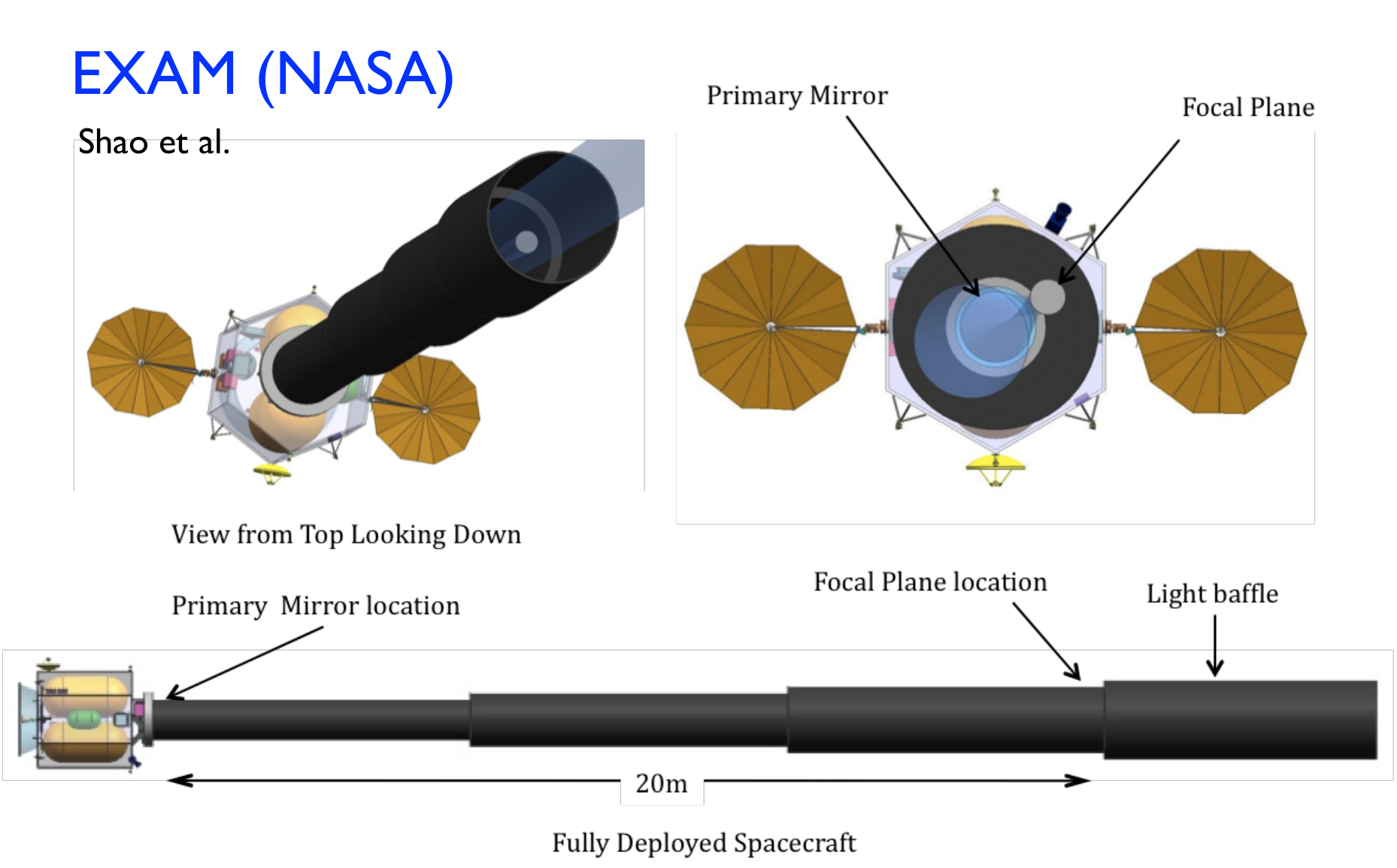}
  \hfill
  \includegraphics[width=0.45\hsize]{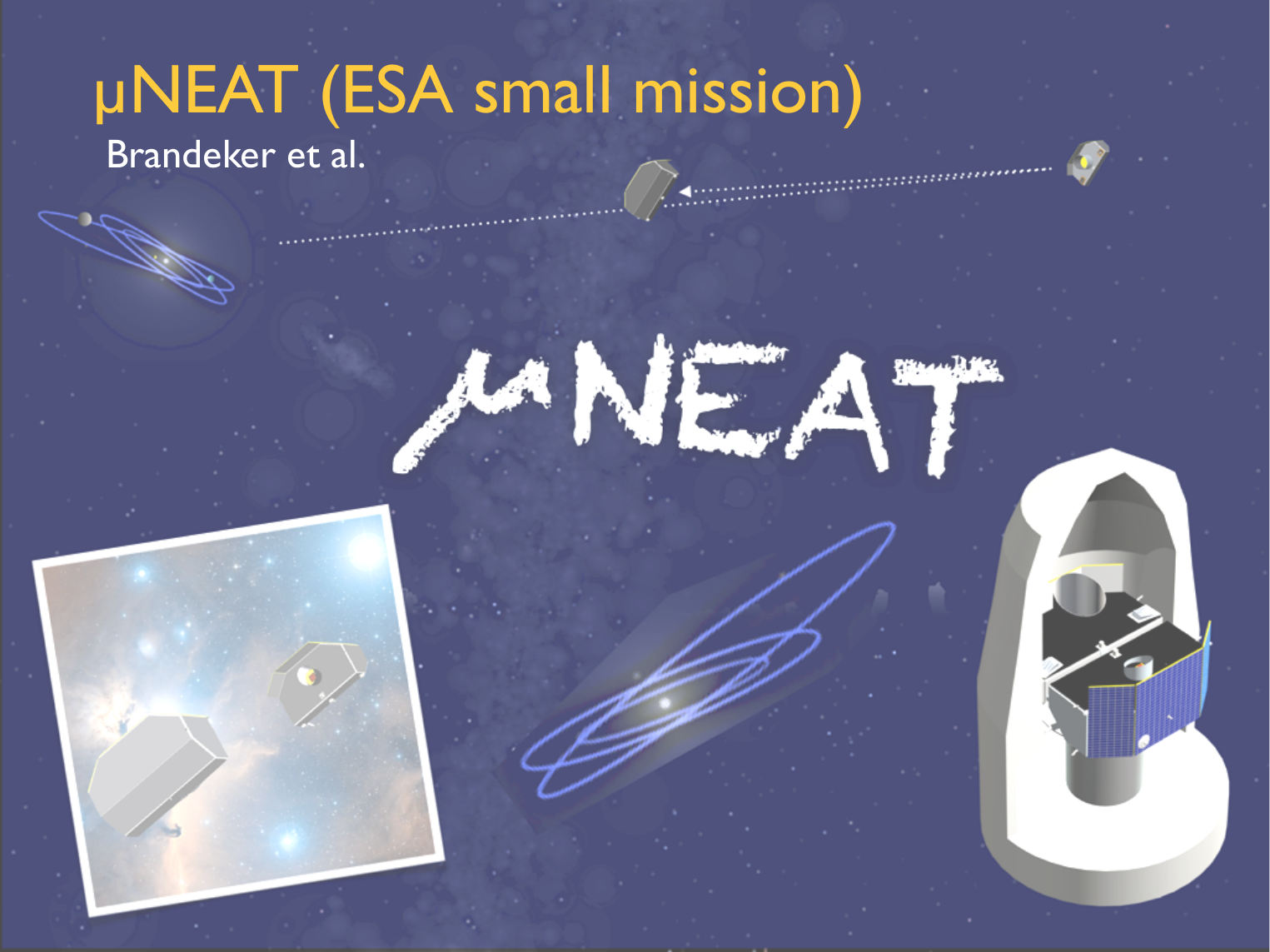}
  \caption{Alternative missions. Left: EXAM, a flight system concept for the deployable telescopic
    tube version for a 0.6\,m version of NEAT. Right: $\mu$NEAT, a
    formation flying version of NEAT with a 0.3\, mirror.}
  \label{fig:alt-concept}
\end{figure}

\textbf{Alternative mission concepts}. An alternative mission concept
would consist of a single spacecraft with an ADAM-like\footnote{ADAM:
  ABLE Deployable Articulated Mast} deployable boom (from ATK-Able
engineering) that connects the telescope and the focal plane
modules. The preliminary investigation made by CNES identified no
show-stoppers for this option: no prohibitive oscillation modes during
observation; during maneuvers, the boom oscillation modes can be
excited but they can be filtered by Kalman filters like SRTM
demonstrated it. The use of dampers on the boom structure allows
damping at a level of 10\% of the oscillations. The main worry
concerns retargeting, which requires large reaction wheels or control
momentum gyroscopes (CMGs) on the spacecraft due to the important
inertia but propellers could be added at the boom end. A possible
implementation made by JPL is shown in left part of
Fig.~\ref{fig:alt-concept}. Concerning a formation flying version, a smaller version of NEAT called $\mu$NEAT
(see right part of Fig.~\ref{fig:alt-concept})
has been recently submitted using existing technology (PRISMA and
state-of-the-art metrology).

\section{Discussion}
\label{sec:discussion}

\subsection{Astrophysical issues}
\label{sec:astro-challenges}

\textbf{Stellar activity.} If all instrumental problems are controlled
then the next obstacle to achieve the scientific objective is of
astrophysical nature, the impact of stellar activity. Spots and bright
structures on the stellar surface induce astrometric, photometric and
RV signals. Using the Sun as a proxy, Lagrange et al.\cite{2011A&A...528L...9L} have
computed the astrometric, photometric and RV variations that would be
measured from an observer located 10 pc away. It appears that the
astrometric variations due to spots and bright structures are small
compared to the signal of an Earth mass planet in the HZ\cite{2010A&A...519A..66M, 2009ApJ...707L..73M,
  2010ApJ...717.1202M}. This remains true throughout the entire solar
cycle. If we consider a star \emph{5 times more active than the active
  Sun}, an Earth-mass planet would still be detectable even during the
highest activity phases. Such activity, or lower, translates in terms
of activity index $\log(R'_{HK}) \leq -4.35$. Consequently, in our
target list, we have kept only stars with such an index (\emph{only
  4\% were discarded}), for which their intrinsic activity should not
prevent the detection of an Earth-mass planet, even during its high
activity period.

\textbf{Perturbations from reference stars.} The vast majority of the
reference stars will be K giants at a distance of $\approx
1$\,kpc. The important parameters in addition to the position are the
proper motion with typical value of $\approx 1$\,mas/yr and the
parallax whose typical value is $\approx 1$\,mas.  They are to be
compared to the accuracy of the cumulative measurements during a
visit. An important value for NEAT accuracy is what is obtained for an
$R=6$ magnitude target: $0.8\,\uas/\sqrt{h}$. The ratios between that
(required) accuracy and the expected motions of the references
indicates clearly that the latter cannot be considered as fixed. Their
positions are members of the set of parameters that have to be solved
for. Because the reference stars are much more distant
($\approx1$\,kpc) than the target star ($\approx10$\,pc), we are 100
times less sensitive to their planetary perturbations. Only
Saturn-Jupiter mass objects matter, and statistically, they are only
present around $\approx10$\% of stars. These massive planets can be
searched for by fitting first the reference star system ($\approx
100\,N_{\rm ref}$ measurements for $5\,N_{\rm ref}$ parameters when
there are no giant planets around the reference stars), possibly
eliminate those with giant planets, and studying the target star with
respect to that new reference frame. Moreover, the largest disturbers
will be detected from ground based radial velocity measurements, and
the early release of Gaia data around 2016 will greatly improve the
position accuracy of the reference stars. For smaller planets at or
below the threshold of detection, their impact on the target
astrometry will be only at a level $\ll 1\,\MEarth$ around
it. Similarly the activity of these K giants has been investigated and
neither the stellar pulsations nor the stellar spots will disturb the
signal at the expected accuracy.

\textbf{Planetary system extraction from astrometric data.} We
recently carried out a major numerical simulation to test how well a
space astrometry mission could detect planets in multi-planet systems
\cite{2010EAS....42..191T}.  The simulation engaged 5 teams of
theorists who generated model systems, and 5 teams of double-blind
\emph{``observers''} who analyzed the simulated data with noise
included.  The parameters of the study were the same as for NEAT,
viz., astrometric single-measurement uncertainty (0.80\,\uas noise,
0.05\,\uas floor, 5-year mission, plus RV observations with 1\,m/s
accuracy for 15\,years).  We found that terrestrial-mass,
habitable-zone planets ($\approx$ Earths) were detected with about the
same efficiency whether they were alone in the system or if there were
several other giant-mass, long-period planets ($\approx$ Jupiters)
present. The reason for this result is that signals with unique
frequencies are well separated from each other, with little
cross-talk.  The number of planets per system ranged from 1 to 11,
with a median of 3. The SNR value of 5.8 value was
predicted\cite{1982ApJ...263..835S} for a false alarm probability
(FAP) of less than 1\%, and verified in our simulations. The
completeness and reliability to detect planets was better than 90\%
for all planets, where the comparison is with those planets that
should have been detected according to a Cramer-Rao estimate
\cite{2010ApJ...720.1073G} of the mission noise.  The Cramer-Rao
estimates of uncertainty in the parameters of mass, semi-major axis,
inclination, and eccentricity were consistent with the “observed”
estimates of each: 3\% for planet mass, $\approx4^{\circ}$ for
inclination and 0.02 for eccentricity.

\textbf{Radial velocity screening.} To solve
unambiguously for giant planets with periods longer than 5\,yrs, it is
necessary to have a ground RV survey for 15\,yrs of the 200 selected
target star, at the presently available accuracy of 1\,m/s. More than
80\% of our targets are already being observed by RV, but the
observations of the rest of them should start soon, well before the
whole NEAT data is available.  The capability of ground based RV
surveys, despite their impressive near-term potential to obtain
accuracies better than 1\,m/s, is not sufficient to detect terrestrial
planets in the HZ of F, G and K stars.  Formally, an accuracy
of 0.05\,m/s is required to see an edge-on Earth mass planet at 1\,\AU
from a solar-mass star with $\SNR=5$, which
might be achievable instrumentally, but is stopped in most cases by
the impact of stellar activity on RV accuracy. It is necessary to find
particularly ``quiet'' stars, but they are a minority (few percents)
and cannot provide a full sample. Furthermore, the ambiguity in
physical mass associated with the signal coming only from the radial
component of the stellar reflex motion ($\sin i$ ambiguity) requires
additional information to determine the physical mass and relative
inclination in complex planetary systems. In some, but not all cases,
limits are possible, and one can argue statistically that 90\% of
systems should be oriented such that the physical planet mass is
within a factor of two of the mass found in RV. However, for finding a
small number of potential future targets for direct detection and
spectroscopy, an absolute determination that the mass is Earth-like is
required as well as an exhaustive inventory of the planets around
stars in our neighborhood.

\begin{table}[t]
  \centering
 \caption{Science impact of NEAT scaling. The nominal mission is
    highlighted in yellow.}
  \label{tab:scaling}
  \includegraphics[width=0.9\hsize]{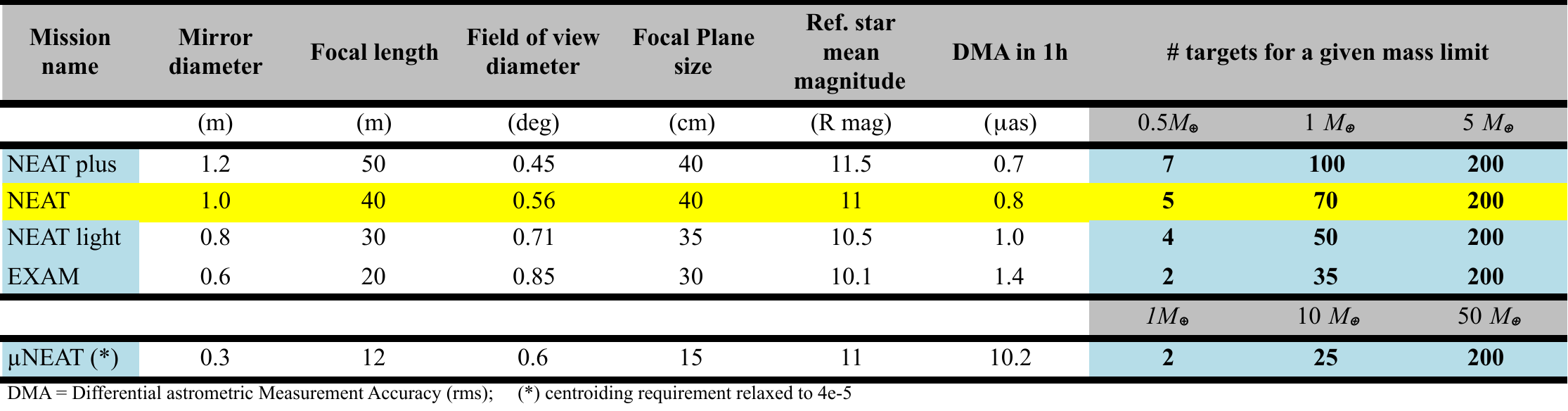}
\end{table}
\textbf{Flexibility of objectives to upgrades / downgrades of the
  mission.} One of the strengths of NEAT is its flexibility, the
possibility to adjust the size of the instrument with impacts on the
science that are not prohibitive. The size of the NEAT mission could
be reduced (or increased) with a direct impact on the accessible
number of targets but not in an abrupt way. For instance, for same
amount of integration time and number of maneuvers, the options listed
in Table \ref{tab:scaling} are possible, with impacts on the number of
stars that can be investigated down to 0.5 and 1 Earth mass, and on
the mass of the instrument, required fuel for maneuvers, and therefore
cost. The time necessary to achieve a given precision depends on the
mass limit that we want to reach: going from 0.5\,\MEarth to
1\,\MEarth requires twice less precision and therefore 4 times less
observing time allowing a smaller telescope. There is room for
adjustment keeping in mind that one wants to survey the neighborhood
with the smallest mass limit possible and a typical number of targets
of $\approx 200$.

\subsection{Technical issues}
\label{sec:technical-challenges}

\textbf{Optical aberrations.} NEAT uses a very simple telescope
optical design. A 1-m diameter clear aperture off-axis parabola, with
an off-axis distance of 1\,m and a 40\,m focal length. The focal plane
is at the prime focus. The telescope is diffraction limited at the
center of the field, where the target stars will be observed, but coma
produces some field dependent aberrations. At the mean position of the
reference stars, $0.2^{\circ}$ away from the center of the field, the
coma produces a \emph{steady} 23\% increase of the point spread
function (PSF) width and an 8\,\microns centroid offset. The impact
remains low since we are looking at differential effects.

\textbf{Centroid measurements.} They consist of two steps: the
determination of the stellar centroid on each CCD during 57\,s and
then the calibration of the relative position of the CCDs during 3\,s
thanks to the metrology. The metrology determines also the response
map of the detectors. As in the normal approach to precision
astrometry with CCDs, we perform a least-square fit of a template PSF
to the pixelated data. PSF knowledge error leads to systematic errors
in the conventional centroid estimation. An accurate centroid
estimation algorithm by reconstructing the PSF from well sampled
(above Nyquist frequency) pixelated images was
developed\cite{2011RSPSA.467.3550Z, 2011SPIE.8151E..28N}. In the limit
of an ideal focal plane array whose pixels have identical response
function (no inter-pixel variation), this method can estimate centroid
displacement between two 32x32 images to sub-micropixel
accuracy. Inter-pixel response variations exist in real CCDs, which we
calibrate by measuring the pixel response of each pixel in Fourier
space
Capturing
inter-pixel variations of pixel response to the third order terms in
the power series expansion, we have shown with simulated data\cite{2011RSPSA.467.3550Z} that the
centroid displacement estimation is accurate to a few micro-pixels.

\textbf{Stability of the primary mirror.} The primary optic will be
made of zerodur/ULE with a temperature coefficient better than
$10^{-8}$/K with an optics thickness $\approx 10$\,cm and
the effective temperature and temperature gradients are kept stable to
$\approx 0.1$\,K over the mirror, the optic is then stable to
$\approx0.1$\,nm ($\lambda/6000$) during the 5\,yr
mission. We have simulated two images, one at the center of the field
that is a perfect Airy function and one at the edge of the field that
has a $\lambda/20$ coma. We added also wavefront errors with a
conservative rms value of $\lambda/1000$. With the new wavefronts, we
calculated the change in the differential astrometry bias caused by
both pixelation and changing wavefronts. While the wavefront
deviations to optimal shape caused a centroid shift of
$\approx6-10\,\uas$ ($10^{-4}$ pixels), differential errors remained
less than $\approx0.3\,\uas$ ($3\times10^{-6}$ pixels).

\textbf{CCD damage in L2 environment.} CCDs suffer damage in radiation
environments\footnote{\url{http://www.rssd.esa.int/gaia}}. \emph{Charge Transfer Efficiency}
(CTE)\footnote{The Gaia community speaks of the complementary
  quantity, charge transfer inefficiency (CTI), in order to emphasize
  its detrimental effects.}, caused by
solar wind protons colliding with the CCD silicon lattice and causing
displacement damage, leads to the formation of traps
which can capture photo-electrons and release them again after some
time. This results in loss of signal and a distortion of the shape of
the PSF image, leading to systematic errors in the image
location due to a mismatch between the ideal PSF shape and the actual
image shape. There are differences between
NEAT and Gaia which justify the assumption that radiation damage
effects will play a much smaller role: NEAT is looking for
extended periods at very bright stars compared to Gaia and is not
operated in time-delayed integration mode. The CCDs are also
regularly illuminated by the laser light from the metrology
system. Therefore the signal level of the pixels
is high, which will keep the traps with long release time constants
filled and effectively inactive.

\textbf{CCD/metrology tests in the lab.} In the absence of optical
errors, the major error sources are associated with the focal plane:
(1) motions of the CCD pixels, which have to be monitored to
$3\times10^{-6}$ pixels every 60\,s, i.e.\ 0.03\,nm; (2) measurements
of the centroid of the star images with $5\times10^{-6}$ pixel
accuracy. We have set up technology testbeds to demonstrate that we
can achieve these objectives. The technology objective for (1) has
almost been reached and the technology demonstration for (2) is
underway and should be completed soon\cite{2012SPIE.8442.....N,
  2012SPIE.8445.....C}. Latest results with no metrology nor QE
6-parameter calibration have been obtained from the CCD / metrology
test bench with performance better $4\times10^{-5}$ pixel at 100s
integration time. The data collected by Nemati et
al. \cite{2011SPIE.8151E..28N} shows that we are only a factor 10 from
the final goal and that differential metrology at intervals of minutes
is required to reach it.

\section{Perspectives}
\label{sec:perspectives}

In the Cosmic Vision plan for 2015-2025, the community has identified
in Theme 1 the question: \emph{``What are the conditions for planet
  formation?''}, and the recommendation in Sect.~1.2: \emph{``Search
  for planets around stars other than the Sun...''} ultra high precise
astrometry as a key technique to explore our solar-like neighbors.
\begin{quote} \em
  ``On a longer timescale, a complete census of all Earth-sized
  planets within 100 pc of the Sun would be highly desirable. Building
  on Gaia's expected contribution on larger planets, this could be
  achieved with a high-precision terrestrial planet astrometric
  surveyor.''
\end{quote}
We have designed NEAT to be this astrometric surveyor. In Europe, as
discussed in detail in the conclusions of the conference
\emph{Pathways to Habitable Planets} \cite{2010ASPC..430.....C} and
in the \emph{Blue Dot Team} report, the exoplanet community recognizes
the importance of astrometric searches for terrestrial planets and has
prioritized this search as a key question in the mid-term, i.e. in the
time frame 2015-2022. The ExoPlanet Task Force (ExoPTF) in the US made
a similar statement. Finally the ESA dedicated \emph{ExoPlanetary
  Roadmap Advisory Team} (EPRAT) prioritizes \emph{Astrometric
  Searches for Terrestrial Planets} in the mid term, i.e.\ in the time
frame 2015-2022. Although the Decadal Survey of Astronomy and Astrophysics for
2010-2020 ranked down the SIM-Lite proposal, but placed as number one
priority a program \emph{``to lay the technical and scientific foundation for
a future mission to study nearby Earth-like planets''}.

Because of these recommandations by the community, we believe that
there is a place for a mission like NEAT in future space programs,
that is to say, a mission that is capable of detecting and
characterizing planetary systems orbiting bright stars in the solar
neighborhood that have a planetary architecture like that of our Solar
System or an alternative planetary system partly composed of
Earth-mass planets. These stars visible with the naked eye or simple
binoculars, if found to host Earth-mass planets, will change
humanity's view of the night sky.

\acknowledgments     
 
This work has benefited support from the Centre National des \'Etudes
Spatiales (CNES), the Jet Propulsion Laboratory (JPL), Thales Alenia
Space (TAS), Swedish Space Corporation (SSC) and the Labex
OSUG@2020. AC PhD fellowship is funded by CNES and TAS.


\bibliography{neat_spie2012}   
\bibliographystyle{spiebib}   

\end{document}